# Dynamics of the Currency Composition of Central Bank Reserves

Deborah Gefang
Department of Economics, Leicester University

Stephen G. Hall
Department of Economics, Leicester University, Bank of Greece, and
University of St Andrews

George S. Tavlas[#,*]
Bank of Greece and the Hoover Institution, Stanford University

August 2026

**Abstract**

We examine how macroeconomic and geopolitical developments in the United States, the euro area, and China affect the currency composition of central banks' foreign exchange reserves. Using an unbalanced panel of reserve shares for 53 countries over 1999–2023, we estimate a constrained system of equations that explicitly imposes the adding-up restriction on reserve shares. The results indicate substantial persistence in reserve holdings and significant cross-currency dependence, supporting a system-wide dynamics of reserve composition. In the country fixed-effects specification (1) issuer economic size, (2) uncertainty, (3) sanctions, (4) trade linkages, and (5) issuer credibility are significantly associated with reserve allocation across currencies. With the inclusion of year fixed effects, the persistence and cross-currency dependence remain, while trade linkages and sanctions emerge as the most important determinants of reserve composition. The results highlight the importance of accounting for the compositional nature and interdependence of reserve shares when examining the determinants of global reserve holdings.

*JEL* Classification: C32, E52, E58

# We thank Falk Laser, Alexander Mihailov, and Jan Weidner for generously providing the most up-to-date data and for their helpful discussions. We have also benefited from comments from seminar participants at the 30th International Conference on Macroeconomic Analysis and International Finance held at the University of Crete in Rethymno, Crete, on May 27-30, 2026. We thank Elisavet Bosdelekidou and Maria Monopoli for research assistance.

* Correspondence may be addressed to George Tavlas, Bank of Greece, 21 E. Venizelos Ave., Athens, 102 50, Greece, Tel. no. +30 210 320 2370; Email address: gtavlas@bankofgreece.gr

## 1. Introduction

A large body of literature has examined the determinants of reserve currency composition, but this literature does so almost exclusively through a currency-by-currency lens. Early contributions emphasise macroeconomic fundamentals such as trade invoicing patterns, financial market depth, capital account openness, exchange rate stability, and geopolitical alignment. More recent work extends these determinants to study network effects, portfolio optimisation frictions, and the growing influence of financial sanctions and geopolitical fragmentation (see below).

A common feature, and an important limitation, of the existing literature is that reserve currency shares are typically modelled as if they evolve independently. This assumption is at odds with the construction of reserve currency shares themselves. To illustrate, suppose that a central bank's holdings of different reserve currencies are measured in a common unit, such as Special Drawing Rights (SDRs). Holding exchange rates and the quantities of all other reserve currencies constant, an increase in holdings of currency A raises the total value of the reserve portfolio. As a result, the share of currency A in total reserves increases, while the shares of all other currencies necessarily decline, even though their nominal holdings remain unchanged. This follows directly from the adding-up constraint; by construction, reserve currency shares must sum to 100 percent.

The above example shows the mechanical interdependence of reserve currency shares. In practice, however, the dynamic is likely to be more complex. The factors that lead a central bank to increase its holdings of currency A may simultaneously affect its demand for currency B, either positively or negatively. For example, stronger trade or financial linkages with the issuer of currency A may increase the demand for A while reducing the relative attractiveness of other currencies. Conversely, developments that increase the demand for A may also increase the demand for B if the two currencies provide complementary liquidity or diversification benefits. Of course, these effects need not be confined to any single alternative currency; changes in the demand for currency A may simultaneously affect the demand for currencies B, C, D, and others, depending on the underlying factors driving these changes.

Reserve currency shares are, therefore, inherently interdependent: any change in one component of the portfolio mechanically alters the measured shares of the others. Ignoring this adding-up constraint has two important implications. First, unconstrained systems may generate economically infeasible predictions when used for simulations or counterfactual analysis. For example, an increase in US real GDP could be estimated to raise the share of US dollar (USD) reserves without inducing a corresponding decline in the shares of other reserve currencies. Such an outcome violates the definition of portfolio shares, which must sum to unity. Second, imposing the adding-up constraint can improve econometric efficiency because it incorporates information that is known to hold by construction. By explicitly recognising the interdependence among reserve currency shares, a constrained system exploits cross-equation restrictions and yields more efficient and internally consistent estimates.

This interdependence also raises a broader question that is largely absent from the existing literature: how do developments in the major economies that issue, anchor, or compete for international reserve currencies affect the allocation of reserves across the global central banking system? Rather than considering the determinants of individual currency shares in isolation, this study takes a system-wide perspective and examines how economic and geopolitical developments in the United States, the euro area, and China shape the allocation of official reserves across currencies. These three economies provide a particularly important setting for such an analysis. The United States and the euro area underpin the two dominant reserve currencies, while China has emerged as an increasingly important economic and geopolitical actor, which has sought to promote a greater international role for the Chinese renminbi (CNY) (e.g., Chinn, Frankel, and Ito, 2024; Ito, 2017). Developments in these three economies' economic performance, monetary policies, trade, and geopolitical relationships can, therefore, affect not only the attractiveness of their respective currencies, but also the relative allocation of reserves across the entire portfolio.

To set the stage for this study, Figure 1 presents the evolution of the currency composition of the IMF's official foreign exchange reserves (COFER) from 1995 to 2025, covering the U.S. dollar (USD), euro, Japanese yen (JPY), British pound sterling (GBP), Swiss franc (CHF), Canadian dollar (CAD), Australian dollar (AUD), and the Chinese renminbi (CNY). As evident, there has been a gradual decline in the share of the U.S. dollar, but that currency continues to account for the lion's share of reserves – almost 60 percent in 2025. With the exception of the euro, whose share has fluctuated around 20 percent, the shares of other currencies have been near or below 5 percent during the entire period.

Against this backdrop, this paper moves beyond isolated currency analysis and instead examines reserve composition as a unified system. We analyse the determinants of reserve allocations in an interlinked framework that explicitly accounts for the zero-sum nature of portfolio rebalancing, thereby providing a more coherent account of how global reserve shares evolve in response to economic, financial, and geopolitical forces.

Our contribution is threefold. First, we formulate reserve composition as an interdependent portfolio system and impose the adding-up restrictions directly in estimation. Second, we quantify own-currency persistence and cross-currency reallocation within a common framework covering eight currency categories. Third, by comparing country fixed-effects and two-way fixed-effects specifications, we distinguish associations driven by cross-country variation from those absorbed by common global shocks. The results show that inertia plays an important role in determining a country's foreign exchange reserve composition, alongside trade exposure and sanctions.

The remainder of the paper is structured as follows. Section 2 provides a theoretical framework for understanding reserve currency allocation. Section 3 reviews the recent empirical literature on the determinants of reserve currency composition and discusses the main approaches used to identify the economic and geopolitical factors underlying these decisions. Section 4 presents the empirical analysis, using panel data to examine how economic and geopolitical factors associated with the United States, the euro area, and China affect changes in the currency composition of official reserves, while accounting for the interdependence among reserve currency shares. Section 5 concludes.

## 2. Determinants of International Currency Use: Theoretical Considerations

An international currency fulfills three basic functions in the international monetary system.[1] It serves as a medium of exchange, a unit of account, and a store of value. As a *medium of exchange*, it is used by private agents both in direct exchange and as a vehicle of indirect exchange between two other currencies in foreign trade and international capital transactions. It is also used by official agents as a vehicle for intervention in foreign exchange markets and for balance-of-payments financing. As a *unit of account*, it is used to invoice merchandise trade, to denominate financial transactions, and, by official agents, to define exchange-rate parities. As a *store of value*, it is used by private agents for investing in financial assets. Similarly, official agents hold international currency and financial assets denominated in it as reserve assets. In what follows, we focus on the determinants of the international use of currencies as reserves by central banks.

[1] This section is based, in part, on Tavlas (1991) and Tavlas and Ozeki (1992).

Several sets of (inter-related) conditions are generally shared by national monies that are used as international reserves.

1. *Economic size and network effects.* As measured by a nation's share of global output, economic size drives the network externalities that make a currency attractive as an international reserve (Chinn, Frankel, and Ito, 2024). A large economy typically generates large volumes of exports and imports. As more economic agents use that country's currency for trade, the currency becomes a standard unit of account. This widespread acceptance creates network externalities whereby the currency becomes more useful simply because others are already using it. Consequently, the share of an international currency's share of global reserves often outstrips the issuing country's share of global GDP.[2]
2. *Economic Stability and Issuer Credibility.* For a currency to be used as an international reserve, there needs to be confidence in the value of the currency and in the credibility of the issuer. Relatively high and/or variable rates of inflation add to the costs of using a currency internationally by generating nominal-exchange-rate depreciation and variability. These effects increase the costs of acquiring information and making efficient calculations about the prices bid and offered for tradeable goods and capital assets. Furthermore, inflation increases the costs of holding a currency by eroding its purchasing power and thus debasing the currency as an international store of value and even as a medium of exchange, since international transactions often entail a lapse of time between the initiation and completion of a transaction. Excessive government spending and/or public debt reduce confidence in a currency because they raise fears that a heavily indebted country will resort to the printing press to inflate away its debt or outright default on its debt. Likewise, a country's current-account and net debtor positions are important to the extent that they affect confidence in the currency by increasing solvency risks.

[2] In this connection, Rogoff (2025, 4-5) reported that in 2024 the U.S. economy accounted for about 25 percent of global output (measured at 2024 market exchange rates), whereas the dollar comprised about 60 percent of global reserves.

3. *Inflation variability and vehicle-currency invoicing*. The role played by prices in disseminating information helps explain why trade in primary products and capital assets is usually denominated in a common vehicle currency. Primary products and capital assets are typically characterized by low levels of product differentiation, and they are traded in competitive markets. Compared with the prices of differentiated manufactured products, prices of primary products and financial assets are subject to more frequent changes. In such markets, using as a numeraire a currency with which most participants are familiar minimizes the costs of information and calculation. Consequently, the choice of an invoicing currency in competitive markets usually narrows to a single currency (or, at most, several vehicle currencies), since it is more efficient to transmit price-change information about homogeneous products in a single currency than through many currencies. For example, if five currencies are used to invoice the price of a primary commodity, there are ten bilateral exchange rates and 30 triangular cross rates to consider.[3] Therefore, in markets for homogeneous products, the economies of communication lead to price quotations being provided in a single currency or, at most, several currencies. These factors help explain why the prices of primary products are typically denominated in U.S. dollars.

4. *Financial markets.* A country should possess financial markets that are substantially free of controls; broad, in that they contain a large assortment of financial instruments including some of which can serve as safe assets; and deep, in that they have well-developed secondary markets. It should also possess financial institutions that are sophisticated and competitive in offshore financial centers. The presence of financial-market controls increases the costs of transacting in a currency. For example, restrictions on the convertibility of a currency result in higher transfer costs (*e.g.*, a greater likelihood of illiquidity), thereby impeding its use as an international reserve. A country that is free of controls, however, and also possesses broad and deep financial markets is in a position to serve as an international banking center. Specifically, it can be expected to provide a high degree of efficiency in international liquidity transformation by accepting short-term, liquid liabilities denominated in its own currency while making long-term, less liquid loans abroad. Related to that, the issuing country needs to have established a rule of law that is deemed to be fair to foreign creditors (Rogoff, 2025, 8).

[3] Likewise, n currencies give rise to n(n-l)/2 bilateral exchange rates and to n(n-l)(n-2)/2 triangular cross rates.

5. *Political stability and military power.* A prerequisite for the use of a national money as an international reserve is confidence in the political stability of the issuing country; political instability generates uncertainty, and undermines the use of a currency as a reserve asset. With regard to military power, Mundell (1983, 189) explained that "it is a fact of historical tradition that the top currency is provided by the top power."[4]

6. *Inertia and Path Dependence:* A hallmark of the literature on the determinants of the international use of reserves is the "extremely persistent" nature of reserve shares. Eichengreen, Chiţu, and Mehl (2016) highlight that, historically, network effects create a winner-take-all environment where the established dominant currency maintains its position long after its underlying economic dominance might have shifted.

7. *Interaction effects.* The structure of the U.S. economy provides an important boost to the international role of the dollar because of the particular circumstance of its economy. As mentioned, an economy's large size drives network externalities. Thus, the fact that the U.S. economy accounts for about 25 percent of global GDP helps underpin the international role of the dollar. At the same time, the U.S. economy is relatively closed to foreign trade.[5] Even though U.S. exports of goods and services account for about 10 percent of global exports, the sum of U.S. exports and imports of goods and services account for only about 24 percent of U.S. GDP, which is a much smaller share than those of other large economies. Hence, the U.S. economy is more insulated from external shocks than other large economies.

## 3. Recent Studies

Until the past 10 years or so, empirical studies of the currency composition of international reserves by-and-large used data contained in the IMF's COFER (Currency Composition of Official Foreign Exchange Reserves) data base. COFER contains aggregate reserves' data.

---

[4] Similarly, Eichengreen (2011, 6) argued: "it is a country's position as a great [economic and military] power that results in the international status of its currency."

[5] Nevertheless, because of the very large size of the U.S. economy, its small share of foreign trade translates into a large share of global trade.

Recent studies have supplemented the COFER data with data drawn from holdings of reserves in individual countries. In what follows, we summarize the main findings of recent studies.[6]

Eichengreen, Chiţu, and Mehl (2016) provided a long-term perspective of the determinants of reserves, using data from IMF historical studies, the IMF's annual reports, and the COFER data base. The authors' basic specification related the holdings of reserves to inertia, proxied by a lagged dependent variable, issuing country size, proxied by the issuer's share of global GDP, and the issuer's credibility, proxied by the appreciation of its currency (under the presumption that exchange rate appreciation increases credibility because it makes holding a currency more attractive and encourages its international use – and vice versa for exchange rate depreciation).[7] The authors took account of 11 currencies that were at one point or another reported as reserve units during a period spanning 1947 to 2013. The authors' estimation was based on random country-effects to account for unobserved country specific variation. The authors found that inertia and the credibility of policies became stronger in the post-Bretton Woods (i.e., post-1973) period, while variables associated with network effects, proxied by the issuer's share of global GDP, under the presumption that an increase in the issuer's share increases the value of the currency as more agents adopt it, weakened. Eichengreen, Chiţu, and Mehl (2016) also found that capital account openness can encourage a currency's use as a reserve.

In a follow-up study, Eichengreen, Mehl, and Chiţu (2018, chap. 7) studied the composition of reserves for 11 currencies during the period 1947 to 2014.[8] The results were similar to those reported in Eichengreen, Chiţu, and Mehl (2016), with the effects of inertia and credibility (i.e., currency appreciation) having strengthened around the time of the breakdown of Bretton Woods. The authors concluded that the strengthening of inertia "may be seen as favoring the leading currency (the dollar), a fact underscored by the resilience of its share in global reserves since the financial crisis of 2008-09" (Eichengreen, Mehl, and Chiţu, 2018, 133). Most other variables used in the Eichengreen, Chiţu, and Mehl (2016) and the Eichengreen, Mehl, and Chiţu (2018) studies were found to have no effect on reserve composition and/or were of the wrong (*a priori*) sign.

---

[6] Brief reviews of the pre-2016 literature have been provided by Iancu et al. (2022) and Laser and Weidner (2022).
[7] The notion that credibility can be proxied by a currency's appreciation was motivated by Devereux and Shi (2013).
[8] As noted, Eichengreen, Chiţu, and Mehl (2016) covered the period 1947 to 2013.

Ito and McCauley (2020) used a panel data set of 58 countries (13 advanced and 45 emerging and developing economies) over the period 1999 to 2017 to examine the determinants of reserves comprised of both U.S. dollars and euros. The authors collected annual data from central banks' annual reports, financial statements and other publicly available information, to construct their data set, allowing them to observe reserve management over time for individual countries.[9] For estimation, Ito and McCauley used six non-overlapping three-year averages of both the dependent and independent variables to account for the variability of some of their variables. The authors found that trade invoicing – that is, the shares of dollars and euros, respectively, in export invoicing – and a currency's co-movements with the dollar -- are key determinants (with equal force) of the dollar's and euro's share of reserves. In a series of robustness checks, Ito and MacCauley found that persistence (as measured with a lagged dependent variable) and the U.S. dollar's share of foreign exchange turnover (used as a proxy for financial market size) positively and significantly affected the dollar's share of reserves. The authors found no evidence that the Great Financial Crisis (2009-11) or the subsequent euro debt crisis (2009-17) caused a structural shift in share of reserves.

Laser and Weidner (2022) used a country-specific dataset of 36 countries plus the euro area to study the determinants of the currency composition of reserves during the period 1996 to 2016. Using panel Tobit estimation on four currencies -- the U.S. dollar, the euro, the pound, and the yen -- Laser and Weidner found that currency pegs and trade patterns were significant determinants of currency composition. In contrast to Ito and McCauley (2020) -- see above -- Laser and Weidner also found that the euro crisis -- proxied by a dummy variable from 2011 until 2016 -- had a negative effect on the share of euros in reserves.

Iancu et al. (2022) investigated the drivers of international currency use at both the global and the individual country levels of the U.S. dollar, the euro, the pound, and the yen, using data from (1) the IMF COPER data base, (2) the data provided in Eichengreen, Chiţu, and Mehl (2016), and (3) individual country data which Iancu et al. (2022) collected from a select group of central banks from 57 economies. The economies comprised both Advanced Economies (AEs) and Emerging Markets and Developing Economies (EMDEs). Using fixed effects estimation, the authors found that over the estimation period (1999-2018), lagged currency share (inertia), currency appreciation (credibility), and financial links (proxied by either the share of a country's external public debt or cross-border bank claims denominated in reserve currency) were important drivers of aggregate reserve shares. The size of the reserve issuer's economy was "not robustly significant across specifications" (Iancu et al., 2022: 890). Trade links or network effects (captured by the trade share of a country with the reserve issuer) were important for the AEs, but were not a robust driver of aggregate reserve shares. Using disaggregated data, Iancu et al. (2022) found that trade links with reserve issuers generally failed to explain reserve shares.

[9] For the Latin American central banks, the authors used data provided by the Latin American Reserve Fund.

Arslanalp, Eichengreen, and Simpson-Bell (2022) studied the determinants of reserve shares of four currencies: the U.S. dollar, the euro, the yen, and the pound sterling using a country-level data set comprising an unbalanced panel for the period 1999 to 2020. The unbalanced panel included data from 80 central banks. The data were based on those compiled by Ito and McCauley (2020) as well as the authors' compilations. Using Seemingly Unrelated Regression (SUR) estimation to account for dependence among the errors in regressions for the four currencies, the authors found that exchange rate pegs against the dollar increase the dollar's shares of reserves (by about 10 percentage points). Other variables yielded mixed results (see Arslanalp, Eichengreen, and Simpson-Bell, 2022, Table 3, 11). For example, variables representing the shares of external debt denominated in each of the four currencies considered yielded results that were "inconsistent, both in sign and significance, presumably because these values are collinear with one another" (2022, 11). The difficulty the authors had in finding a cohesive and significant set of results for currency shares has permeated the literature on the determinants of those shares. Typically, authors have been able to identify only a few variables that have an *a priori* expected and significant effect on reserve shares.

Chinn, Frankel, and Ito (2024), studied the determinants of the currency composition of reserves on a currency-by-currency basis for five currencies (the U.S. dollar, euro, the Chinese renminbi, the pound sterling, and the Japanese yen) over the period 1999-2022. Using the Ito and McCauley (2020) data, the authors found that economic size and inertia, along with a few other variables, were significant factors determining the shares of those currencies in reserves. Those authors also pooled their reserves data, imposing the constraint that reserve holdings are determined in the same way for each currency, and found that economic determinants (e.g. economic size, foreign exchange turnover, bilateral currency pegs, and bilateral trade shares) were significant. In a follow-up paper, Chinn, Frankel, and Ito (2025) examined the effects of financial sanctions, trade policy uncertainty, and U.S. tariffs on holding of U.S. dollar reserves; the authors found that the former two variables had a significant effect on dollar reserves while U.S. tariffs did not have a statistically significant impact.

Goldberg and Hannaoui (2026) investigated the determinants of the U.S. dollar's share of reserves during the period 1999 to 2023. The authors applied a decomposition that showed that two distinct channels have contributed to the changes in the dollar's share of reserves aggregated across countries: shifts in preferences for dollar assets and changes in reserve balances driven by countries whose portfolio allocations differ from the aggregate. The authors also found that in the later years of their sample the key contributors to changes in aggregate preference shifts for dollar assets were changes in bilateral country trade with the United States and the dollar's debt share, with geopolitics additionally working through the investment tranches of central bank portfolios.

To sum up, the recent literature expands on the COFER data base, using enlarged data sets (drawn largely from country-specific sources) so that demand for reserve share by individual countries can be studied. By-and-large, this literature has found that inertia, credibility of the issuing country, and the U.S. share of global GDP are drivers of the U.S. dollar's share of reserves. The results with respect to other determinants of reserves have, in general, been inconclusive.

## 4. Panel Data Analyses

IMF COFER data are available only for the world aggregate and a limited set of country groups, rather than for individual countries. To obtain country-level reserve shares, we use the unbalanced panel data set of Laser, Mihailov, and Weidner (2024), which provides currency shares for 64 countries/regions over the period 1999 -- 2023. As mentioned, similar data sets are compiled by Arslanalp, Eichengreen, and Simpson-Bell (2022), Iancu et al. (2022), and Ito and McCauley (2020). We rely on the former because it provides more recent coverage, extending the sample through 2023 and thus capturing both the COVID-19 period and the onset of Russia's invasion of Ukraine.

The reserve currency categories in Laser, Mihailov, and Weidner (2024) consist of the Australian dollar (AUD), Canadian dollar (CND), Chinese renminbi (CNY), euro (EUR), Japanese yen (JPY), pound sterling (GBP), the U.S. dollar (USD), and an 'Other' category that aggregates all remaining reserve currencies (such as the Swiss franc). Excluding the major reserve currency issuer countries and euro area countries from the year these countries adopted the euro onward, our sample comprises 53 countries/regions.

Figures 2.1–2.6 plot the composition of reserve currencies at the country level. In contrast to the aggregate trends presented in Figure 1, the country-level evidence reveals substantial heterogeneity, with considerable variation in both the timing and the magnitude of changes in reserve currency composition. Across the majority of non-European economies, the USD emerges as the dominant reserve currency, with its share remaining persistently high or increasing over time. This pattern is particularly evident across many African, Asian, and Latin American countries. In contrast, the EUR displays strong geographical concentration, maintaining a dominant role in European and EU-adjacent economies such as Bosnia, Bulgaria, Croatia, Lithuania, Slovakia, and Slovenia. While the EUR remains stable or gains importance within Europe, its relative significance generally declines in many non-European economies.

Beyond the two dominant currencies, sterling generally trends downwards, while the renminbi remains a small reserve currency in most countries despite increases in some recent observations. The Australian dollar, Canadian dollar, and yen typically occupy peripheral positions. The country panels also reveal abrupt shifts and missing observations, underscoring both the heterogeneity of reserve management and the unbalanced nature of the sample.

Following previous studies (discussed in Section 3), we include measures of inertia, the issuing country's share of world GDP, the bilateral trade share of the reserve-holding country with the issuing country relative to the former's total trade with the world, credibility (appreciation of the reserve issuer's currency), and geopolitical variables such as sanctions exposure, as determinants of the reserve currency shares held by individual countries. We also include the inflation rate of the reserve-issuing country relative to the average inflation rate of OECD countries as a proxy for monetary policy credibility, and the difference between the issuing country's World Uncertainty Index (WUI) and the average WUI across the G7 countries.[10] Finally, following the existing literature, we include the volatility of the reserve currency to account for exchange rate risk.

GDP, inflation and bilateral trade data are sourced from the IMF, while sanctions data are taken from Yalcin et al. (2025), The Global Sanctions Data Base—Release 4. The exchange rate of the reserve currency is measured as the number of SDR per unit of the currency. Using daily exchange rate data from the IMF, we construct annual average exchange rates and compute appreciation rate as the year-on-year percentage change in these annual averages. Exchange rate volatility is proxied by the within-year standard deviation of daily exchange rate observations.

[10] The WUI developed by Ahir, Bloom, and Furceri (2022) is constructed from references to economic and political uncertainty in Economist Intelligence Unit (EIU) country reports using a standardized methodology and reporting structure, making it particularly suitable for cross-country analysis. By contrast, alternative uncertainty measures, such as the VIX, are largely US-centric and therefore less appropriate as measures of uncertainty for other reserve-currency-issuing countries.

Our constrained panel data model takes the following form:

$$\text{share}_{ijt} = \beta_{0,ij} + \sum_{k=1}^{7} \beta_{1i,h}\, share_{kj,t-1} + \sum_{h=1}^{3} \beta_{2i,h}\, gdp_share_{ht} + \sum_{h=1}^{3} \beta_{3i,h}\, wui_diff_{ht} + \sum_{h=1}^{3} \beta_{4i,h}\, trade_share_{hjt} + \sum_{h=1}^{3} \beta_{5i,h}\, INF_diff_{ht} + \sum_{h=1}^{3} \beta_{6i,h}\, sanction_{hjt} + \sum_{h=1}^{3} \beta_{7i,h}\, Appreciation_{ht} + \sum_{h=1}^{3} \beta_{8i,h}\, Volatility_{ht} + u_{ijt} \tag{1}$$

where $share_{ijt}$ is the share of reserves denominated in currency $i$, ($i = 1, \dots, 8$), held by country $j$ at time $t$; $gdp_share_{ht}$ denotes the share of world GDP accounted for by issuing economy $h$ at time $t$, where $h = 1,2,3$ corresponds to the United States, euro area, and China, respectively. $wui_{ht}$ is the world uncertainty index of reserve issuing country $h$ at time $t$; $trade_{share_{hjt}}$ measures bilateral trade between country $h$ and $j$ as a share of country $j$'s total trade with the rest of the world; $INF_{diff_{ht}}$ is the inflation difference of reserve issuing country $h$ and that of the OECD inflation rate; $sanction_{hjt}$ denotes sanction imposed by country $h$ on country $j$,[11] and $Appreciation_{ht}$ *and* $Volatility_{ht}$ are the appreciation rate and volatility of the reserve currency issued by country $h$. As mentioned, currency appreciation has been used to capture credibility.

As argued in the introduction, for each country, holding foreign exchange reserves, an increase (or decrease) in the share of one reserve currency necessarily implies adjustments in the composition of all other reserve currencies due to the adding-up constraint. Expressing the reserve share of each currency using model (2), we impose adding-up constraints on the system of eight equations where $i$ can be either AUD, CAD, CNY, EUR, JPY, GBP, USD, and Other:

$$\sum_{i=1}^{8} \beta_{0,ij} = 100\,,\ \sum_{i=1}^{8} \beta_{1i,h} = 0,\ \dots,\ \sum_{i=1}^{8} \beta_{8i,h} = 0 \tag{2}$$

[11] We construct the sanctions variable as a dummy indicator, taking the value of 0 when no sanctions are imposed and 1 otherwise. In the empirical analysis, we also distinguish between trade sanctions and financial sanctions. The results are qualitatively similar to those obtained using the aggregate sanctions dummy and are therefore not reported separately.

These restrictions reflect the compositional nature of reserve currency shares. In particular, ceteris paribus, the intercept terms across the eight equations must sum to 100 (in percentage), while the coefficients associated with each explanatory variable must sum to zero across equations. Consequently, any increase in the share of one reserve currency induced by changes in the explanatory variables must be offset by corresponding declines in the shares of other reserve currencies, ensuring that total reserve shares continue to sum to 100%.

We first estimate the constrained Seemingly Unrelated Regression (SUR) system using country-demeaned data. This transformation removes time-invariant country heterogeneity and is equivalent to a one-way fixed-effects specification, while SUR allows contemporaneous correlation across currency equations. The coefficient restrictions impose the adding-up identity on every regressor.

Over the 25-year period spanned by the data, the global economy has experienced several major episodes of turbulence, including financial crises, the Great Recession, the COVID-19 pandemic, and Russia's invasion of Ukraine. To properly account for these common global shocks alongside cross-country heterogeneity, we next use two-way transformed (unit- and time-demeaned) data - removing both country and time fixed effects - and re-estimate the SUR system subject to the constraints described in equations (2).[12]

### 4.1 Empirical Results for One-way Demeaned Data

Table 1 reports the constrained SUR results for the eight reserve currencies using one-way transformed data.[13] Several features of the results are noteworthy. First, the lagged reserve shares are statistically significant for most currencies, indicating substantial persistence in the currency composition of official reserves. The coefficients are particularly large for the own-lagged shares of the AUD, CAD, CNY, JPY, GBP, and USD.[14] This persistence is consistent with the relatively gradual adjustment of central banks' reserve portfolios and suggests that existing reserve allocations are an important determinant of subsequent allocations. However, ranging from 0.5802 for the AUD to 0.8291 for the CNY, these coefficients are generally lower than those reported in the existing literature, which are typically around 0.9. This difference may reflect the inclusion of country–currency pair fixed effects in our specification, which account for unobserved time-invariant characteristics that may otherwise contribute to the estimated persistence. Among the currencies, the CNY has the largest own-lag coefficient, suggesting that its share in central bank reserves is relatively persistent and adjusts more slowly.

[12] An alternative approach to imposing the adding-up restriction is to model the log-ratios of reserve currency shares. Although this approach is straightforward to estimate, its results are sensitive to the choice of benchmark currency and can be difficult to interpret due to the nonlinear nature of the model specification. We relegate the relevant discussion to Online Appendix B.

[13] As a robustness check, we estimate an alternative four-share specification comprising USD, EUR, CNY, and Other, which together sum to 100%. The results, reported in Table A2 of the Online Appendix, are broadly consistent with the eight-share model.

[14] The lagged share of the EUR is excluded from the system as it is highly correlated with the lagged share of USD, with a correlation coefficient of -0.8869.

At the same time, the lagged shares of other currencies are also statistically significant in several equations and generally enter with negative coefficients. This pattern is consistent with the interdependence implied by the adding-up constraint. Changes in the share of one reserve currency are associated with changes in the shares of other currencies, rather than each currency evolving independently.

The GDP shares of China and the euro area are highly correlated, with a correlation coefficient of -0.9388; we therefore dropped the former from the system. Unsurprisingly, a higher US share of world GDP is positively associated with the US dollar share, while it is negatively associated with several other reserve currencies. Similarly, the euro area's share of world GDP is positively and significantly associated with the euro share. In both cases, the estimated coefficients indicate that a 1 percentage-point increase in the issuing economy's share of world GDP is associated with an increase of around 0.45 percentage points in the share of reserves denominated in its currency. These results suggest that the economic size of a reserve-currency issuer is an important factor in determining reserve currency shares.

The uncertainty measure (WUI) for the United States generally does not have significant effects on other reserve currency shares. For the US dollar, however, the coefficient is negative and statistically significant at -0.0840. This suggests that higher relative uncertainty in the United States is associated with a lower share of US dollar reserves. This finding is particularly relevant given the heightened uncertainty surrounding the global economic and geopolitical environment in recent years.

The WUI for the euro area shows a somewhat stronger pattern. It is positively and significantly associated with the Australian dollar share (0.0262), while it is negatively and significantly associated with the euro share (-0.2010). This indicates that an increase in relative uncertainty in the euro area is associated with a reduction in the euro's share of reserves and, at the same time, an increase in the share allocated to the Australian dollar. The coefficients on the remaining currencies are statistically insignificant. Overall, these results suggest that greater uncertainty in the euro area may lead to some reallocation away from the euro towards alternative reserve currencies, although the magnitude of the effect on the Australian dollar is relatively small.

In contrast, none of the coefficients on the China WUI is statistically significant. This suggests that uncertainty in China does not have a statistically discernible effect on the allocation of official reserves across the currencies considered in the system.

The impact of trade exposure to the United States, euro area, and China varies across reserve currencies. Trade exposure to the United States is positively and significantly associated with the euro share, with a coefficient of 0.3388, while it is negatively and significantly associated with the share of 'Other' currencies, with a coefficient of -0.1991. The coefficients on the remaining currency shares are statistically insignificant. The positive association between trade exposure to the United States and the euro share is somewhat unexpected and may reflect factors beyond bilateral trade itself, such as the invoicing currency used in international trade or the broader financial and institutional links between the reserve-holding country and the major economies.

Trade exposure to China shows a different pattern. It is negatively and significantly associated with the euro share (-0.3418) and the pound share (-0.0989), while it is positively and significantly associated with the US dollar share (0.2871) and the share of other currencies (0.1551). The positive association between trade exposure to China and the US dollar share may be consistent with the continued importance of the US dollar as a dominant invoicing and settlement currency in international trade, including trade involving China. The coefficients on the remaining currencies are statistically insignificant. In contrast, none of the coefficients on trade exposure to the euro area is statistically significant.

Monetary policy credibility appears to matter, although the evidence is limited. The US inflation differential is negatively and significantly associated with the Australian dollar share (-0.2135) and positively and significantly associated with the 'Other' currency share (1.1276), while the remaining coefficients are statistically insignificant. The euro area inflation differential is positively and significantly associated with the Japanese yen share (0.5265), but is insignificant in the other equations. The coefficients on the China inflation differential are statistically insignificant across all reserve currencies.

The sanctions results are particularly striking. US sanctions are positively and significantly associated with the CNY share, with a coefficient of 0.3412, while the coefficients for the other currencies are statistically insignificant. EU sanctions show an even more pronounced pattern: they are positively and significantly associated with the CNY share (0.5495) and the "Other" share (5.2108), but negatively and significantly associated with the EUR share (−3.5680) and the JPY share (-2.4739). Finally, Chinese sanctions are positively and significantly associated with the JPY share (1.2898), while the coefficients for the other currencies are statistically insignificant. Overall, these results provide evidence that sanctions are associated with substantial shifts in the composition of official reserves, with the direction and magnitude of these shifts varying considerably across both the sanctioning economy and the reserve currency.

U.S. economic and political credibility (USD appreciation) is positively and significantly associated with that country's USD reserve share (0.6444), while it is negatively and significantly associated with the CNY (-0.0980) and EUR (-0.8618) shares. Similarly, CNY appreciation is positively and significantly associated with the CNY share (0.0925), although it has no significant effect on the shares of the other currencies. Exchange-rate volatility produces a more limited and mixed pattern. US dollar volatility is positively and significantly associated with the EUR share (2.8567) and negatively and significantly associated with the US dollar share (-2.7186). In contrast, neither EUR nor CNY volatility has a statistically significant association with any reserve currency share.

Taken together, the results support the view that reserve currency allocation is a system of interdependent decisions rather than a collection of independent currency choices. The significant cross-currency effects, together with the persistence in reserve shares, indicate that changes in the economic and geopolitical conditions of major reserve-currency issuers can influence the composition of the broader reserve portfolio. At the same time, the substantial heterogeneity across equations suggests that these effects are currency-specific, reinforcing the importance of estimating the reserve-share equations jointly rather than imposing a common response across currencies.

### 4.2 Empirical Results for Two-way Demeaned Data

The results using two-way transformed data are reported in Table 2. [15] Relative to the one-way transformed specification, several notable differences emerge. Most importantly, the economic size variables lose their statistical significance once year effects are removed. Neither the US share nor the euro area share of world GDP is statistically significant in any of the reserve currency equations. In the one-way specification, these variables were among the more important determinants of reserve composition, particularly for the US dollar and the euro. Their loss of significance suggests that the estimated relationships are sensitive to the inclusion of common time effects. This is not surprising, given that the relative economic size of the United States and the euro area varies mainly over time, leaving relatively limited cross-country variation once year effects are removed.

[15] Results for smaller SUR model with four dependent variables are reported in Table A3.

Despite these differences, persistence in reserve holdings remains a central feature of the results. The coefficients on the lagged reserve shares remain highly significant and of similar magnitude to those in the one-way specification. The own-lag coefficients are 0.6794 for the US dollar, 0.8380 for the renminbi, and 0.7612 for the euro, with the renminbi again showing the highest degree of persistence, indicating the importance of inertia even after controlling for country-specific and common time effects.

The uncertainty variables become less important in the two-way specification. None of the US, euro area, or Chinese uncertainty measures is statistically significant in any equation. This contrasts with the one-way results, where euro area uncertainty was negatively associated with the euro share and positively associated with the Australian dollar share. The difference suggests that these relationships are sensitive to the inclusion of common time effects.

Trade linkages, by contrast, remain among the more robust determinants of reserve composition. Trade exposure to the euro area is positively and significantly associated with the euro share (0.6509), while it is negatively associated with the US dollar (-0.1908) and Other (-0.4451) shares. Trade exposure to China also remains negatively associated with the euro share and positively associated with the US dollar share, although its effect on the 'Other' share is no longer statistically significant. The persistence of these trade effects after controlling for year effects suggests that cross-country differences in trade relationships remain an important factor in determining reserve currency compositions.

The inflation differential variables lose their statistical significance in the two-way specification. None of the inflation measures is significant in any of the reserve currency equations, compared with several significant coefficients in the one-way specification. Similarly, exchange-rate appreciation and volatility generally cease to be statistically significant once year effects are included.

Sanctions, however, remain an important exception. US sanctions continue to be positively associated with the euro and renminbi shares and negatively associated with the ‘Other’ share, while EU sanctions are associated with a higher renminbi share and lower euro and ‘Other’ shares.

Overall, the two-way specification produces a somewhat different picture from the one-way model. Several variables that were significant in the baseline specification -- particularly GDP shares, uncertainty, inflation differentials, and exchange-rate variables -- lose significance after common time effects are removed. Trade linkages and sanctions, however, remain significant, while the strong persistence of reserve shares is largely unchanged. These results suggest that differences across countries in their trade relationships and geopolitical exposure provide more robust information about reserve currency allocation than common changes in global economic and financial conditions.

**5. Conclusions**

This paper analyses the currency composition of official foreign exchange reserves as an interdependent portfolio system. Because reserve shares must sum to 100 per cent, a change in one currency’s share necessarily entails changes in the shares of others. Our constrained SUR framework incorporates this adding-up restriction directly and, therefore, provides internally consistent estimates of how economic, financial, and geopolitical developments originating in the United States, the euro area, and China are associated with reserve reallocation across eight currency categories.

Three important findings emerge from the analysis. First, reserve allocations are highly persistent, confirming the importance of inertia and the gradual adjustment of central-bank portfolios. This result is remarkably consistent across both the one-way and two-way fixed-effects specifications. At the same time, the significant cross-currency lag effects indicate that reserve shares do not evolve independently, but rather as components of an integrated portfolio.

Second, the one-way specification suggests that issuer fundamentals matter for reserve currency status. A larger US share of world GDP is associated with a higher dollar reserve share, while a larger euro area share of world GDP is associated with a higher euro reserve share. Similarly, appreciation of the dollar and renminbi, which the literature has interpreted as capturing credibility, is associated with increases in their respective reserve shares, whereas greater dollar volatility is associated with a reallocation away from the dollar and towards the euro. However, these effects become statistically insignificant once time fixed effects are introduced, suggesting that their explanatory power may reflect broad global trends rather than persistent differences across countries.

Third, trade and geopolitical relationships emerge as among the most robust determinants of reserve currency allocation. Trade exposure to China is consistently associated with a higher US dollar share and a lower euro share, while stronger trade links with the euro area are associated with greater euro holdings, even after controlling for common global shocks. Particularly striking are the effects of sanctions. Across both specifications, sanctions imposed by the United States and the European Union are associated with significant reserve reallocations, including increases in renminbi holdings and reductions in the shares of some traditional reserve currencies. By contrast, the effects of uncertainty, inflation differentials, and exchange-rate variables largely disappear in the two-way specification. These results suggest that geopolitical exposure and trade integration provide more robust explanations of reserve portfolio choices than common changes in the global macroeconomic environment.

Overall, the results point to an international reserve system that remains strongly influenced by the United States, while also demonstrating the importance of geopolitical considerations in reserve allocation. The continued association between greater trade exposure to China and higher dollar, rather than renminbi holdings, helps explain the continuing dominance of the dollar in international invoicing, settlement, and financial markets. At the same time, the sanctions' results suggest that geopolitical fragmentation may create incentives for some central banks to diversify their reserve portfolios and increase their holdings of renminbi. Nevertheless, the strong persistence observed in reserve allocations implies that changes in the international currency hierarchy are likely to occur gradually. The future roles of the US dollar, the euro, and the renminbi will therefore depend not only on the economic performance of their issuing economies, but also on how evolving trade relationships and geopolitical tensions reshape the incentives facing reserve-holding central banks.

## Table 1: Currency Composition of Central Bank Reserves (One-Way Demeaned)

| *Variable* | *AUD* | *CAD* | *CNY* | *EUR* | *GBP* | *JPY* | *USD* | *OTHER* |
|---|---|---|---|---|---|---|---|---|
| *AUD.L1* | ***0.5802**** | *0.0078* | *0.0123* | ***-0.6255**** | *0.0562* | *0.0026* | *0.1708* | *-0.2043* |
| | *(0.0274)* | *(0.0201)* | *(0.0218)* | *(0.1541)* | *(0.0524)* | *(0.0322)* | *(0.1773)* | *(0.1313)* |
| *CAD.L1* | ***0.1326**** | ***0.7381**** | ***-0.0860*** | ***-0.6054*** | *0.0798* | *-0.0401* | *-0.1427* | *-0.0763* |
| | *(0.0387)* | *(0.0284)* | *(0.0308)* | *(0.2177)* | *(0.0740)* | *(0.0454)* | *(0.2505)* | *(0.1854)* |
| *CNY.L1* | *0.0067* | *0.0038* | ***0.8291**** | ***-0.4993*** | *0.0603* | *-0.0032* | *-0.1786* | *-0.2188* |
| | *(0.0319)* | *(0.0234)* | *(0.0253)* | *(0.1790)* | *(0.0608)* | *(0.0373)* | *(0.2059)* | *(0.1525)* |
| *GBP.L1* | ***0.0447**** | *0.0051* | *0.0107* | ***-0.4143**** | ***0.7657**** | ***0.0456*** | *-0.0128* | ***-0.4448**** |
| | *(0.0130)* | *(0.0095)* | *(0.0103)* | *(0.0730)* | *(0.0248)* | *(0.0152)* | *(0.0840)* | *(0.0622)* |
| *JPY.L1* | *0.0058* | *0.0107* | *-0.0176* | ***-0.2755*** | *0.0257* | ***0.6673**** | ***-0.3821*** | *-0.0344* |
| | *(0.0219)* | *(0.0161)* | *(0.0174)* | *(0.1229)* | *(0.0418)* | *(0.0256)* | *(0.1414)* | *(0.1047)* |
| *USD.L1* | ***0.0099*** | *0.0009* | *-0.0029* | ***-0.4670**** | *0.0095* | ***0.0128*** | ***0.6514**** | ***-0.2146**** |
| | *(0.0047)* | *(0.0035)* | *(0.0037)* | *(0.0265)* | *(0.0090)* | *(0.0055)* | *(0.0304)* | *(0.0225)* |
| *US_GDP_share* | ***-0.0624**** | ***-0.0599*** | *0.0235* | *-0.3217* | *0.0032* | *-0.0276* | ***0.4870*** | *-0.0421* |
| | *(0.0369)* | *(0.0271)* | *(0.0294)* | *(0.2076)* | *(0.0706)* | *(0.0433)* | *(0.2388)* | *(0.1768)* |
| *EA_GDP_share* | *-0.0030* | *-0.0157* | *-0.0197* | ***0.4423*** | *0.0581* | *-0.0217* | *-0.2246* | *-0.2156* |
| | *(0.0376)* | *(0.0276)* | *(0.0299)* | *(0.2115)* | *(0.0719)* | *(0.0441)* | *(0.2433)* | *(0.1802)* |
| *WUI_USD* | *0.0030* | *0.0034* | *-0.0127* | *0.0345* | *0.0307* | *0.0138* | *-0.0840* | *0.0112* |
| | *(0.0129)* | *(0.0095)* | *(0.0103)* | *(0.0725)* | *(0.0246)* | *(0.0151)* | *(0.0834)* | *(0.0617)* |
| *WUI_Europe* | ***0.0262**** | *0.0054* | *0.0039* | ***-0.2010*** | *0.0089* | *-0.0110* | *0.0630* | *0.1046* |
| | *(0.0153)* | *(0.0112)* | *(0.0122)* | *(0.0861)* | *(0.0293)* | *(0.0180)* | *(0.0991)* | *(0.0734)* |
| *WUI_CN* | *0.0066* | *0.0058* | *-0.0031* | *-0.0104* | *0.0083* | *0.0034* | *-0.0216* | *0.0109* |
| | *(0.0071)* | *(0.0052)* | *(0.0056)* | *(0.0397)* | *(0.0135)* | *(0.0083)* | *(0.0457)* | *(0.0338)* |
| *Trade_Share_with US* | *-0.0135* | *0.0047* | *-0.0039* | ***0.3388*** | *-0.0323* | *-0.0263* | *-0.0684* | ***-0.1991**** |
| | *(0.0252)* | *(0.0185)* | *(0.0201)* | *(0.1417)* | *(0.0482)* | *(0.0296)* | *(0.1631)* | *(0.1207)* |

| *Variable* | *AUD* | *CAD* | *CNY* | *EUR* | *GBP* | *JPY* | *USD* | *OTHER* |
|---|---|---|---|---|---|---|---|---|
| *Trade_share_with _EA* | *0.0005* | *0.0008* | *-0.0226* | *0.0241* | *-0.0116* | *-0.0210* | *0.0506* | *-0.0208* |
| | *(0.0190)* | *(0.0140)* | *(0.0151)* | *(0.1070)* | *(0.0364)* | *(0.0223)* | *(0.1231)* | *(0.0911)* |
| *Trade_share_with_CN* | *0.0099* | *-0.0095* | *0.0024* | ***-0.3418**** | ***-0.0989**** | *-0.0042* | ***0.2871**** | ***0.1551*** |
| | *(0.0120)* | *(0.0088)* | *(0.0096)* | *(0.0675)* | *(0.0229)* | *(0.0141)* | *(0.0777)* | *(0.0575)* |
| *INF_diff_US* | ***-0.2135** | *0.0634* | *0.0493* | *-0.6972* | *0.3240* | *-0.0217* | *-0.6319* | ***1.1276** |
| | *(0.1239)* | *(0.0909)* | *(0.0985)* | *(0.6960)* | *(0.2366)* | *(0.1452)* | *(0.8008)* | *(0.5929)* |
| *INF_diff_EA* | *0.1631* | *0.0300* | *-0.2390* | *0.3279* | *0.2429* | ***0.5265*** | *-0.3749* | *-0.6765* |
| | *(0.2216)* | *(0.1626)* | *(0.1762)* | *(1.2450)* | *(0.4231)* | *(0.2598)* | *(1.4323)* | *(1.0605)* |
| *INF_diff_CN* | *0.0547* | *-0.0068* | *-0.0321* | *-0.0030* | *0.0887* | *0.0661* | *0.0933* | *-0.2609* |
| | *(0.0427)* | *(0.0314)* | *(0.0340)* | *(0.2402)* | *(0.0816)* | *(0.0501)* | *(0.2763)* | *(0.2046)* |
| *US_SANCTION* | *0.2879* | *-0.0229* | ***0.3412*** | *0.7465* | *-0.4947* | *0.1721* | *-0.5031* | *-0.5269* |
| | *(0.1996)* | *(0.1465)* | *(0.1587)* | *(1.1214)* | *(0.3811)* | *(0.2340)* | *(1.2901)* | *(0.9551)* |
| *EU_SANCTION* | *0.0175* | *-0.1134* | ***0.5495*** | ***-3.5680*** | *0.3443* | *0.0332* | *-2.4739* | ***5.2108**** |
| | *(0.2944)* | *(0.2161)* | *(0.2341)* | *(1.6545)* | *(0.5623)* | *(0.3452)* | *(1.9034)* | *(1.4092)* |
| *CN_SANCTION* | *-0.4015* | *-0.4625* | *0.0173* | *-3.9408* | *-0.7500* | ***1.2898*** | *5.0065* | *-0.7589* |
| | *(0.4935)* | *(0.3622)* | *(0.3924)* | *(2.7730)* | *(0.9424)* | *(0.5786)* | *(3.1902)* | *(2.3620)* |
| *USD_appreciation* | *0.0165* | *0.0308* | ***-0.0980*** | ***-0.8618*** | *0.1418* | *0.0080* | ***0.6444** | *0.1182* |
| | *(0.0569)* | *(0.0417)* | *(0.0452)* | *(0.3195)* | *(0.1086)* | *(0.0667)* | *(0.3676)* | *(0.2722)* |
| *CNY_appreciation* | *0.0144* | *-0.0257* | ***0.0925*** | *0.3215* | *-0.0644* | *-0.0096* | *-0.1529* | *-0.1758* |
| | *(0.0526)* | *(0.0386)* | *(0.0418)* | *(0.2955)* | *(0.1004)* | *(0.0617)* | *(0.3399)* | *(0.2517)* |
| *USD_volatility* | *-0.1186* | *-0.0492* | *0.1190* | ***2.8567*** | *0.2137* | *0.1326* | ***-2.7186** | *-0.4357* |
| | *(0.2312)* | *(0.1697)* | *(0.1839)* | *(1.2994)* | *(0.4416)* | *(0.2711)* | *(1.4948)* | *(1.1068)* |
| *EUR_volatility* | *-0.0433* | *0.0161* | *-0.1308* | *-0.6753* | *0.1950* | *-0.0652* | *1.0293* | *-0.3259* |
| | *(0.1472)* | *(0.1080)* | *(0.1170)* | *(0.8271)* | *(0.2811)* | *(0.1726)* | *(0.9515)* | *(0.7045)* |
| *CNY_volatility* | *-0.7998* | *-0.2388* | *-0.3561* | *-1.3545* | *0.5578* | *-0.2615* | *0.7171* | *1.7359* |
| | *(0.7033)* | *(0.5163)* | *(0.5592)* | *(3.9522)* | *(1.3432)* | *(0.8247)* | *(4.5468)* | *(3.3664)* |

| *Variable* | *AUD* | *CAD* | *CNY* | *EUR* | *GBP* | *JPY* | *USD* | *OTHER* |
|---|---|---|---|---|---|---|---|---|
| ***Observations*** | *777* | *777* | *777* | *777* | *777* | *777* | *777* | *777* |
| ***Adjusted R-squared*** | ***0.4801*** | ***0.5290*** | ***0.6436*** | ***0.4813*** | ***0.7033*** | ***0.4866*** | ***0.5296*** | ***0.1780*** |

*Notes: Standard errors are in parentheses below coefficients. Significant coefficients are shown in bold. * $p<0.10$, ** $p<0.05$, *** $p<0.001$.*

**Table 2: Currency Composition of Central Bank Reserves (Two-way Demeaned)**

| *Variable* | *AUD* | *CAD* | *CNY* | *EUR* | *GBP* | *JPY* | *USD* | *OTHER* |
|---|---|---|---|---|---|---|---|---|
| *AUD.L1* | ***0.5785**** | *0.0085* | *0.0106* | ***-0.6493**** | *0.0553* | *0.0002* | *0.1631* | *-0.1669* |
| | *(0.0274)* | *(0.0200)* | *(0.0217)* | *(0.1562)* | *(0.0528)* | *(0.0318)* | *(0.1774)* | *(0.1323)* |
| *CAD.L1* | ***0.1319*** | ***0.7433**** | ***-0.0791*** | ***-0.5350*** | *0.0699* | *-0.0361* | *-0.1359* | *-0.1590* |
| | *(0.0385)* | *(0.0282)* | *(0.0305)* | *(0.2198)* | *(0.0744)* | *(0.0448)* | *(0.2496)* | *(0.1861)* |
| *CNY.L1* | *-0.0017* | *0.0037* | ***0.8310**** | ***-0.5164*** | *0.0571* | *-0.0130* | *-0.1624* | *-0.1984* |
| | *(0.0317)* | *(0.0232)* | *(0.0251)* | *(0.1808)* | *(0.0612)* | *(0.0368)* | *(0.2053)* | *(0.1531)* |
| *GBP.L1* | ***0.0442**** | *0.0059* | *0.0060* | ***-0.4788**** | ***0.7612**** | ***0.0365*** | *0.0126* | ***-0.3876**** |
| | *(0.0126)* | *(0.0092)* | *(0.0100)* | *(0.0717)* | *(0.0242)* | *(0.0146)* | *(0.0814)* | *(0.0607)* |
| *JPY.L1* | *0.0014* | *0.0107* | *-0.0132* | *-0.1701* | *0.0307* | ***0.6767**** | ***-0.4639*** | *-0.0724* |
| | *(0.0221)* | *(0.0161)* | *(0.0175)* | *(0.1259)* | *(0.0426)* | *(0.0256)* | *(0.1430)* | *(0.1066)* |
| *USD.L1* | ***0.0087** | *0.0015* | *-0.0043* | ***-0.4931**** | *0.0083* | *0.0082* | ***0.6583**** | ***-0.1877**** |
| | *(0.0045)* | *(0.0033)* | *(0.0035)* | *(0.0255)* | *(0.0086)* | *(0.0052)* | *(0.0290)* | *(0.0216)* |
| *US_GDP_share* | *0.3913* | *0.0423* | *-0.1450* | *0.6131* | *-0.3727* | *0.0904* | *-0.4794* | *-0.1400* |
| | *(0.4710)* | *(0.3443)* | *(0.3734)* | *(2.6871)* | *(0.9092)* | *(0.5473)* | *(3.0517)* | *(2.2758)* |
| *EA_GDP_share* | *-0.3396* | *-0.1695* | *0.0943* | *-0.8457* | *-0.2706* | *-0.3784* | *2.0918* | *-0.1823* |
| | *(0.4276)* | *(0.3125)* | *(0.3389)* | *(2.4391)* | *(0.8253)* | *(0.4968)* | *(2.7701)* | *(2.0657)* |
| *WUI_USD* | *0.0157* | *-0.0118* | *-0.0472* | *0.3119* | *0.1275* | *-0.0888* | *-0.6613* | *0.3539* |
| | *(0.1686)* | *(0.1232)* | *(0.1336)* | *(0.9616)* | *(0.3254)* | *(0.1959)* | *(1.0921)* | *(0.8144)* |
| *WUI_Europe* | ***-0.4730** | *0.1245* | *-0.0991* | *-0.2011* | *0.2630* | *0.1140* | *0.6092* | *-0.3375* |
| | *(0.2613)* | *(0.1910)* | *(0.2072)* | *(1.4909)* | *(0.5044)* | *(0.3037)* | *(1.6931)* | *(1.2626)* |
| *WUI_CN* | *-0.0512* | *0.0404* | *-0.1074* | *-0.1116* | *-0.1755* | *-0.0595* | *0.6069* | *-0.1421* |
| | *(0.1199)* | *(0.0877)* | *(0.0951)* | *(0.6842)* | *(0.2315)* | *(0.1394)* | *(0.7770)* | *(0.5795)* |
| *Trade_Share_with US* | *-0.0085* | *0.0092* | *-0.0074* | ***0.2705** | *-0.0333* | *-0.0172* | *-0.0341* | *-0.1792* |
| | *(0.0247)* | *(0.0180)* | *(0.0196)* | *(0.1408)* | *(0.0477)* | *(0.0287)* | *(0.1599)* | *(0.1193)* |

| *Variable* | *AUD* | *CAD* | *CNY* | *EUR* | *GBP* | *JPY* | *USD* | *OTHER* |
|---|---|---|---|---|---|---|---|---|
| *Trade_share_with_EA* | *0.0063* | *0.0032* | *-0.0129* | ***0.4775**** | *-0.0195* | *0.0078* | ***-0.1975*** | ***-0.2649**** |
| | *(0.0120)* | *(0.0088)* | *(0.0095)* | *(0.0684)* | *(0.0231)* | *(0.0139)* | *(0.0777)* | *(0.0579)* |
| *Trade_share_with_CN* | *0.0122* | *-0.0100* | *0.0029* | ***-0.3182**** | ***-0.0981**** | *-0.0104* | ***0.2763**** | ***0.1453*** |
| | *(0.0117)* | *(0.0085)* | *(0.0092)* | *(0.0665)* | *(0.0225)* | *(0.0136)* | *(0.0756)* | *(0.0564)* |
| *INF_diff_US* | *-0.4898* | *-0.1611* | *-0.0911* | *1.4372* | *1.3957* | *-1.1231* | *-2.6099* | *1.6421* |
| | *(1.9934)* | *(1.4568)* | *(1.5801)* | *(11.3713)* | *(3.8475)* | *(2.3161)* | *(12.9142)* | *(9.6305)* |
| *INF_diff_EA* | ***8.3963** | *0.3279* | *0.2332* | *7.4934* | *-0.8260* | *2.4909* | *-20.5829* | *2.4673* |
| | *(4.6567)* | *(3.4032)* | *(3.6913)* | *(26.5646)* | *(8.9881)* | *(5.4106)* | *(30.1689)* | *(22.4978)* |
| *INF_diff_CN* | *0.8934* | *0.1843* | *-0.3389* | *1.4832* | *-0.4475* | *0.0061* | *-1.5278* | *-0.2529* |
| | *(0.9041)* | *(0.6607)* | *(0.7166)* | *(5.1573)* | *(1.7450)* | *(1.0504)* | *(5.8570)* | *(4.3677)* |
| *US_SANCTION* | *0.3239* | *-0.0147* | ***0.3555*** | *0.8288* | *-0.5966* | *0.2031* | *-0.4144* | *-0.6857* |
| | *(0.2000)* | *(0.1461)* | *(0.1585)* | *(1.1407)* | *(0.3859)* | *(0.2323)* | *(1.2954)* | *(0.9660)* |
| *EU_SANCTION* | *0.0375* | *-0.1355* | ***0.5340*** | ***-3.0010** | *0.3454* | *0.0499* | *-3.0900* | ***5.2597**** |
| | *(0.2925)* | *(0.2138)* | *(0.2319)* | *(1.6689)* | *(0.5647)* | *(0.3399)* | *(1.8953)* | *(1.4134)* |
| *CN_SANCTION* | *-0.3208* | *-0.4158* | *0.0285* | *-3.0042* | *-0.8455* | ***1.3565*** | *4.7413* | *-1.5400* |
| | *(0.4930)* | *(0.3603)* | *(0.3908)* | *(2.8126)* | *(0.9516)* | *(0.5729)* | *(3.1942)* | *(2.3820)* |
| *USD_appreciation* | ***1.3897** | *-0.1996* | *0.3406* | *-2.2321* | *-0.1989* | *-0.8534* | *0.7097* | *1.0440* |
| | *(0.8041)* | *(0.5877)* | *(0.6374)* | *(4.5871)* | *(1.5520)* | *(0.9343)* | *(5.2095)* | *(3.8849)* |
| *CNY_appreciation* | *-0.0726* | *-0.0121* | *-0.3115* | *2.5857* | *-0.6945* | *0.2961* | *-1.3109* | *-0.4802* |
| | *(0.7629)* | *(0.5576)* | *(0.6048)* | *(4.3522)* | *(1.4726)* | *(0.8864)* | *(4.9427)* | *(3.6859)* |
| *USD_volatility* | *2.9280* | *-0.5912* | *0.3993* | *5.8347* | *1.2304* | *-0.1954* | *-12.2325* | *2.6266* |
| | *(2.5628)* | *(1.8730)* | *(2.0315)* | *(14.6198)* | *(4.9466)* | *(2.9777)* | *(16.6034)* | *(12.3816)* |
| *EUR_volatility* | *-1.1034* | *0.6406* | *0.1378* | *-6.5162* | *2.3040* | *1.0203* | *6.6400* | *-3.1230* |
| | *(2.0364)* | *(1.4883)* | *(1.6143)* | *(11.6171)* | *(3.9306)* | *(2.3662)* | *(13.1934)* | *(9.8386)* |
| *CNY_volatility* | *5.9781* | *-1.3274* | *2.0902* | *40.6602* | *6.0903* | *-6.7745* | *-39.1370* | *-7.5799* |
| | *(14.5041)* | *(10.5999)* | *(11.4974)* | *(82.7404)* | *(27.9952)* | *(16.8525)* | *(93.9668)* | *(70.0736)* |

| *Variable* | *AUD* | *CAD* | *CNY* | *EUR* | *GBP* | *JPY* | *USD* | *OTHER* |
|---|---|---|---|---|---|---|---|---|
| ***Observations*** | *777* | *777* | *777* | *777* | *777* | *777* | *777* | *777* |
| ***Adjusted R-squared*** | ***0.4575*** | ***0.5062*** | ***0.6214*** | ***0.5620*** | ***0.6399*** | ***0.4888*** | ***0.5429*** | ***0.1149*** |

***Notes:*** *Standard errors are in parentheses below coefficients. Significant coefficients are shown in bold. * p<0.10, ** p<0.05, *** p<0.001*

Figure 1: Shares of major currencies in central banks foreign exchange reserves

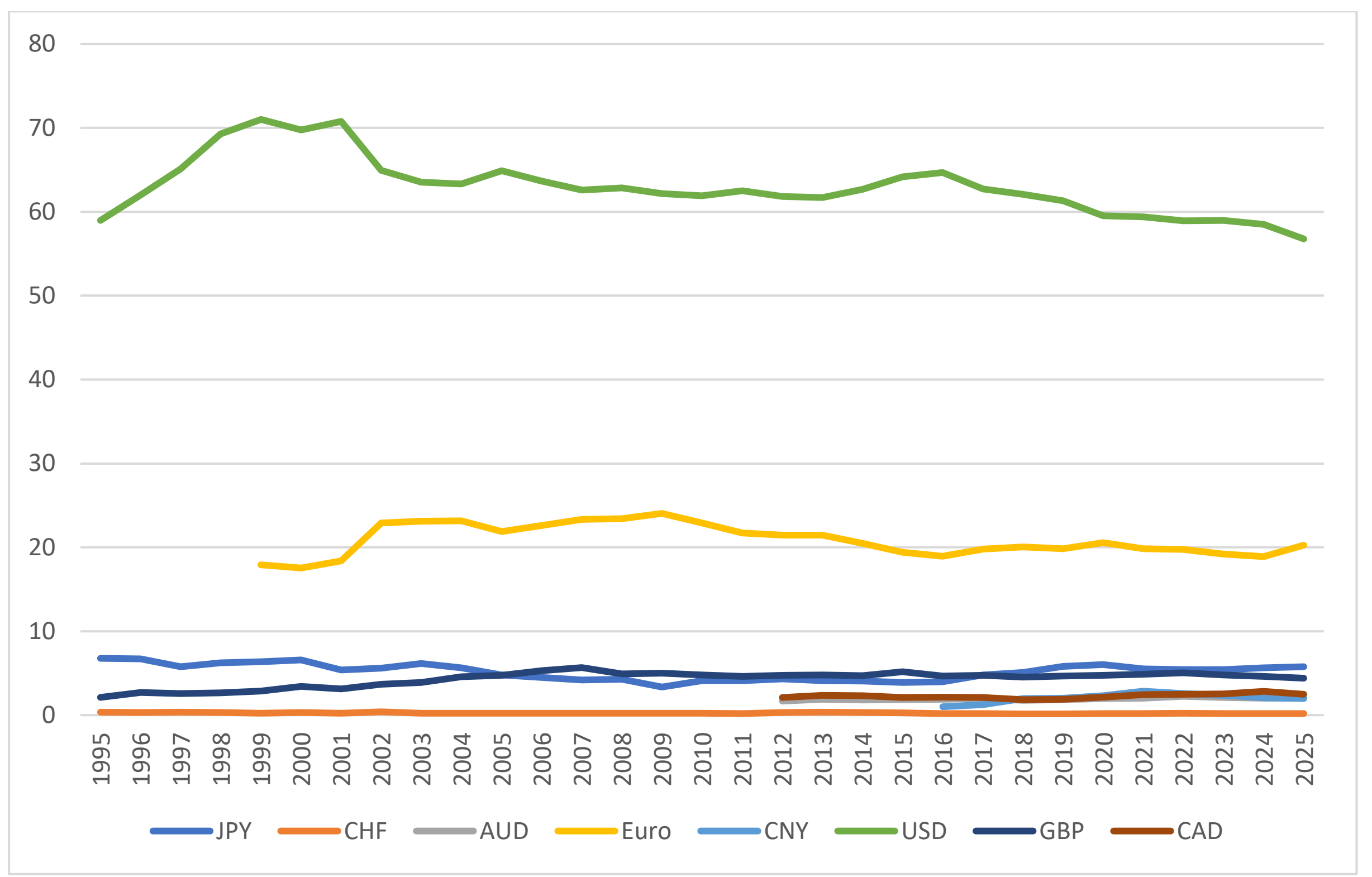

Figure 2.1

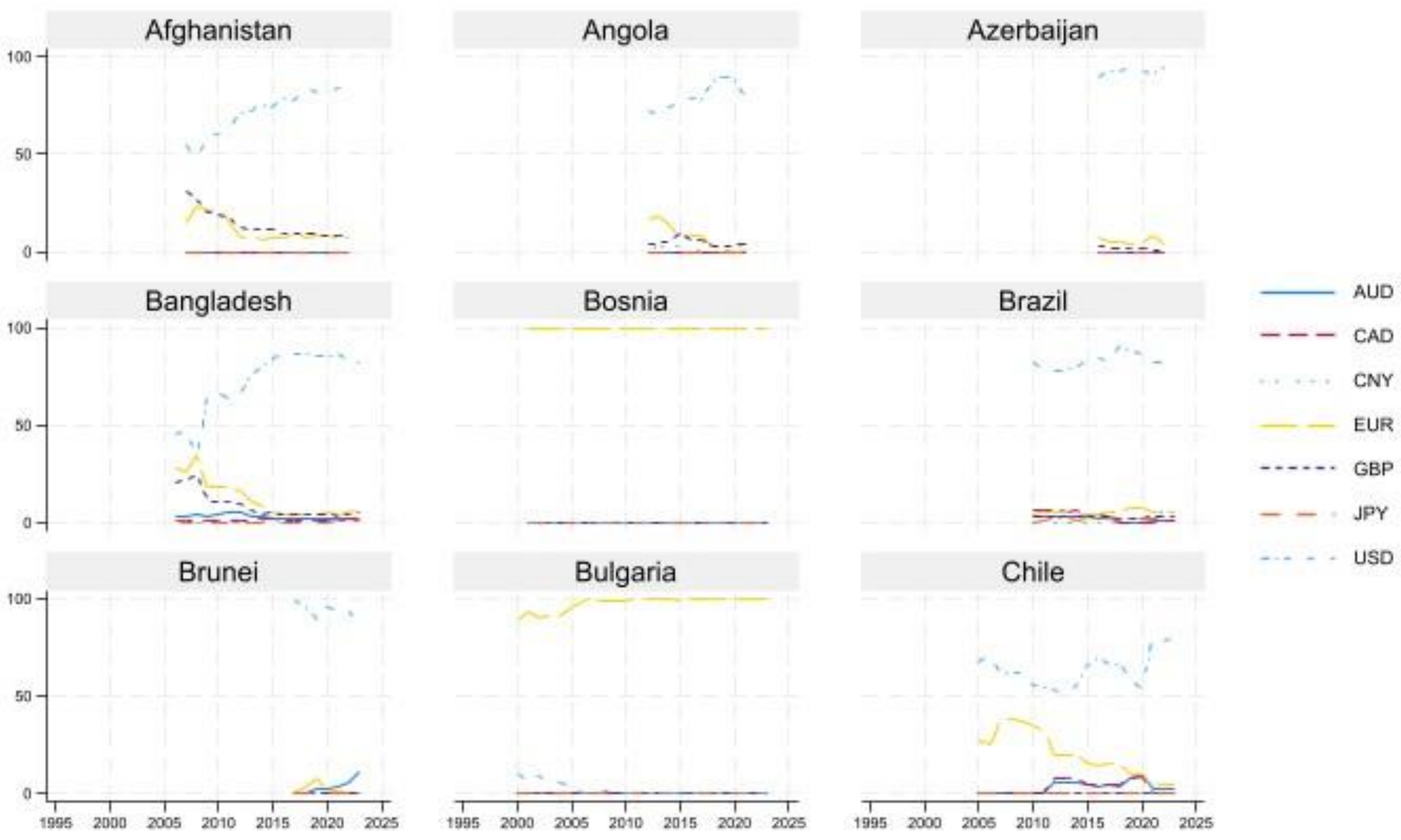
Afghanistan
Angola
Azerbaijan
Bangladesh
Bosnia
Brazil
Brunei
Bulgaria
Chile
AUD
CAD
CNY
EUR
GBP
JPY
USD

Figure 2.2

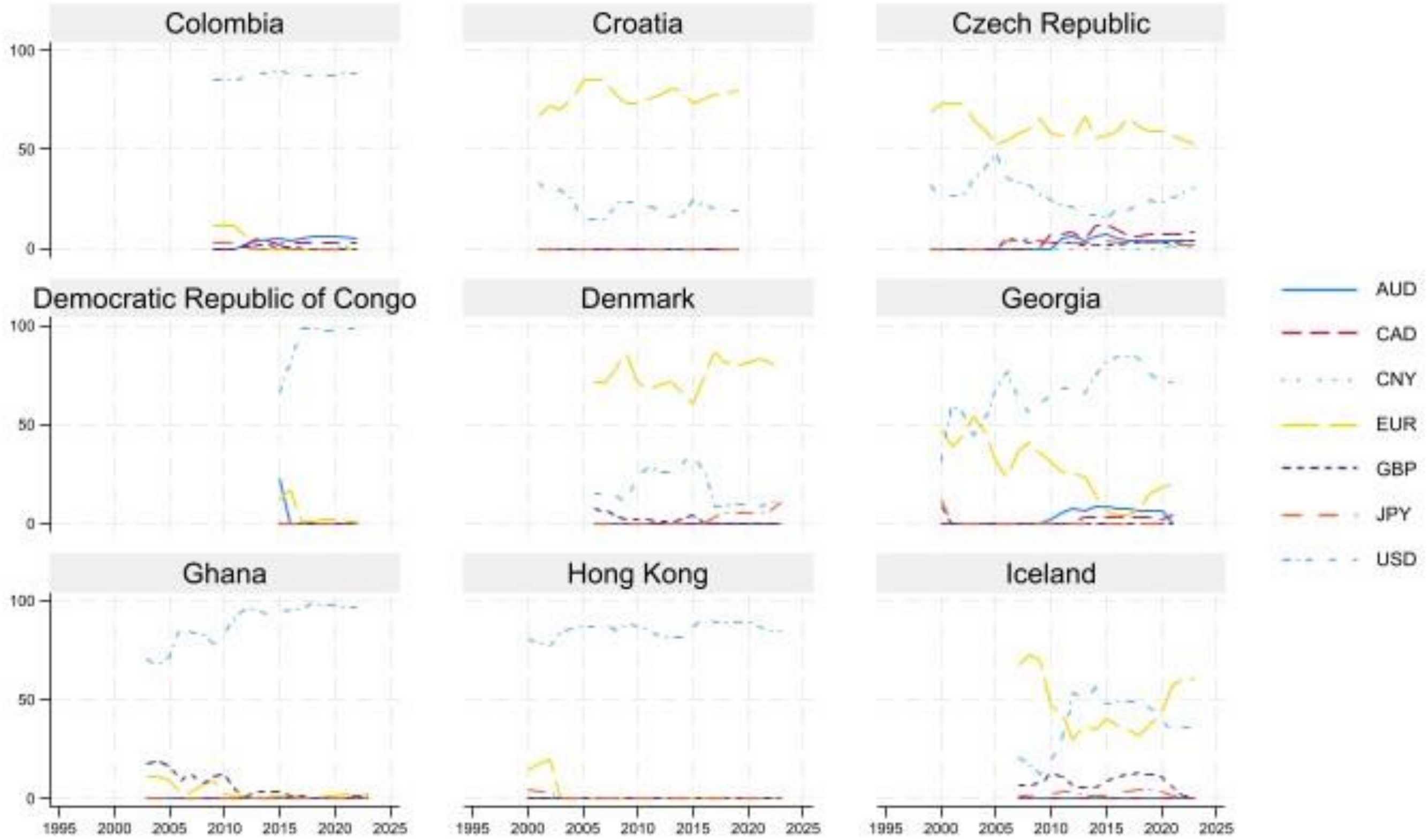
Colombia
Croatia
Czech Republic
Democratic Republic of Congo
Denmark
Georgia
Ghana
Hong Kong
Iceland
100
50
0
1995
2000
2005
2010
2015
2020
2025
AUD
CAD
CNY
EUR
GBP
JPY
USD

Figure 2.3

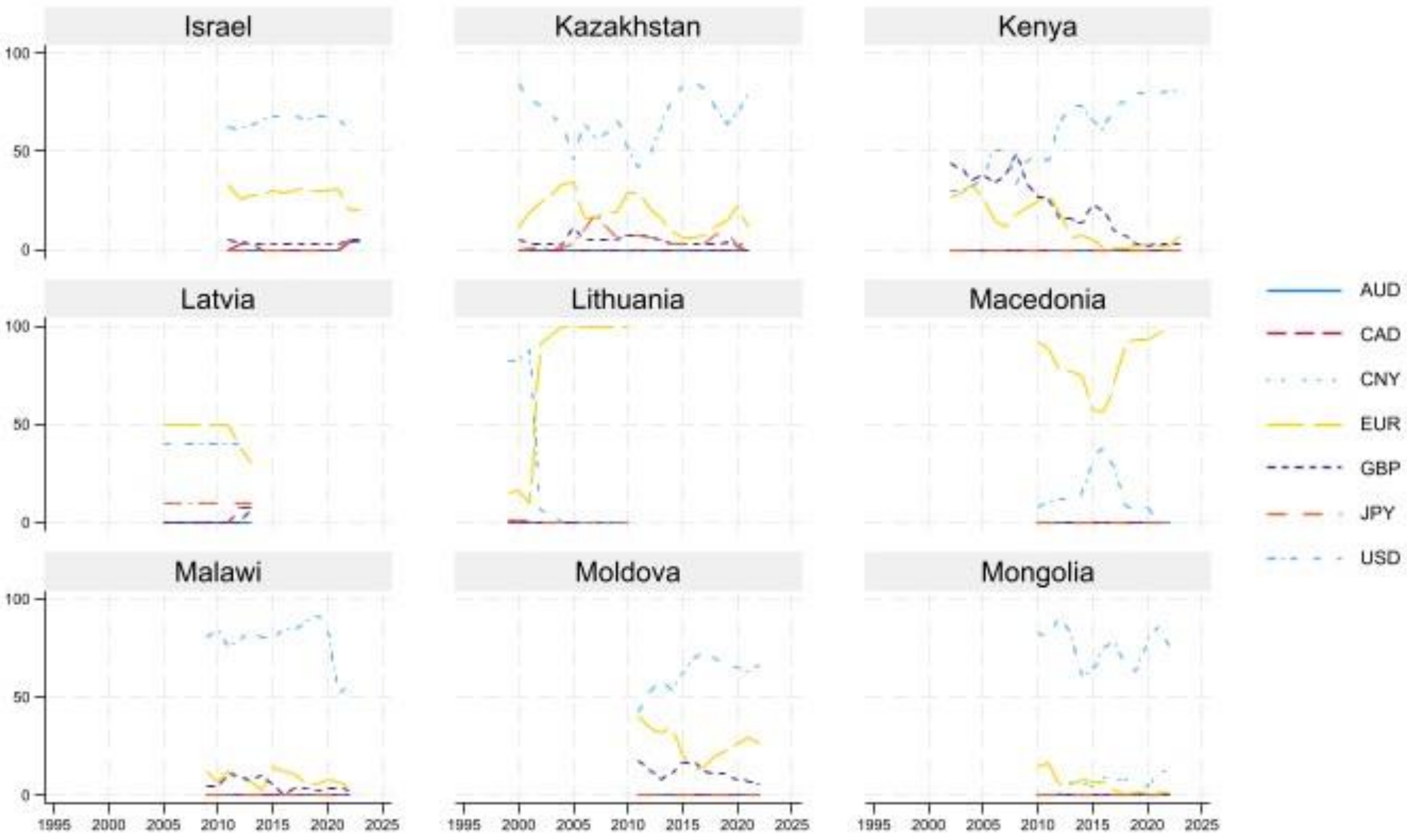
Israel
Kazakhstan
Kenya
Latvia
Lithuania
Macedonia
Malawi
Moldova
Mongolia
100
50
0
1995
2000
2005
2010
2015
2020
2025
AUD
CAD
CNY
EUR
GBP
JPY
USD

Figure 2.4

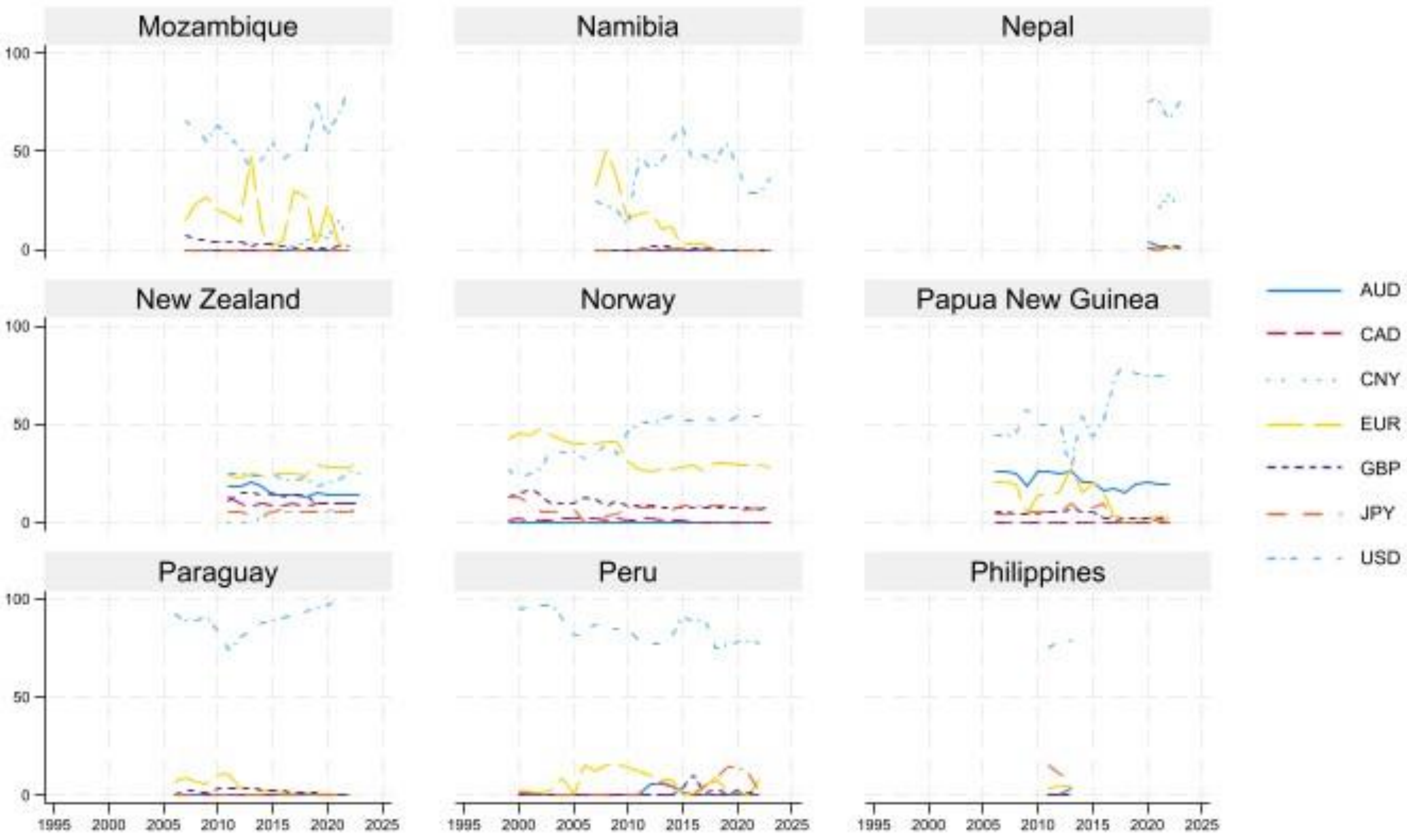

Mozambique
Namibia
Nepal
New Zealand
Norway
Papua New Guinea
Paraguay
Peru
Philippines
100
50
0
1995
2000
2005
2010
2015
2020
2025
AUD
CAD
CNY
EUR
GBP
JPY
USD

Figure 2.5

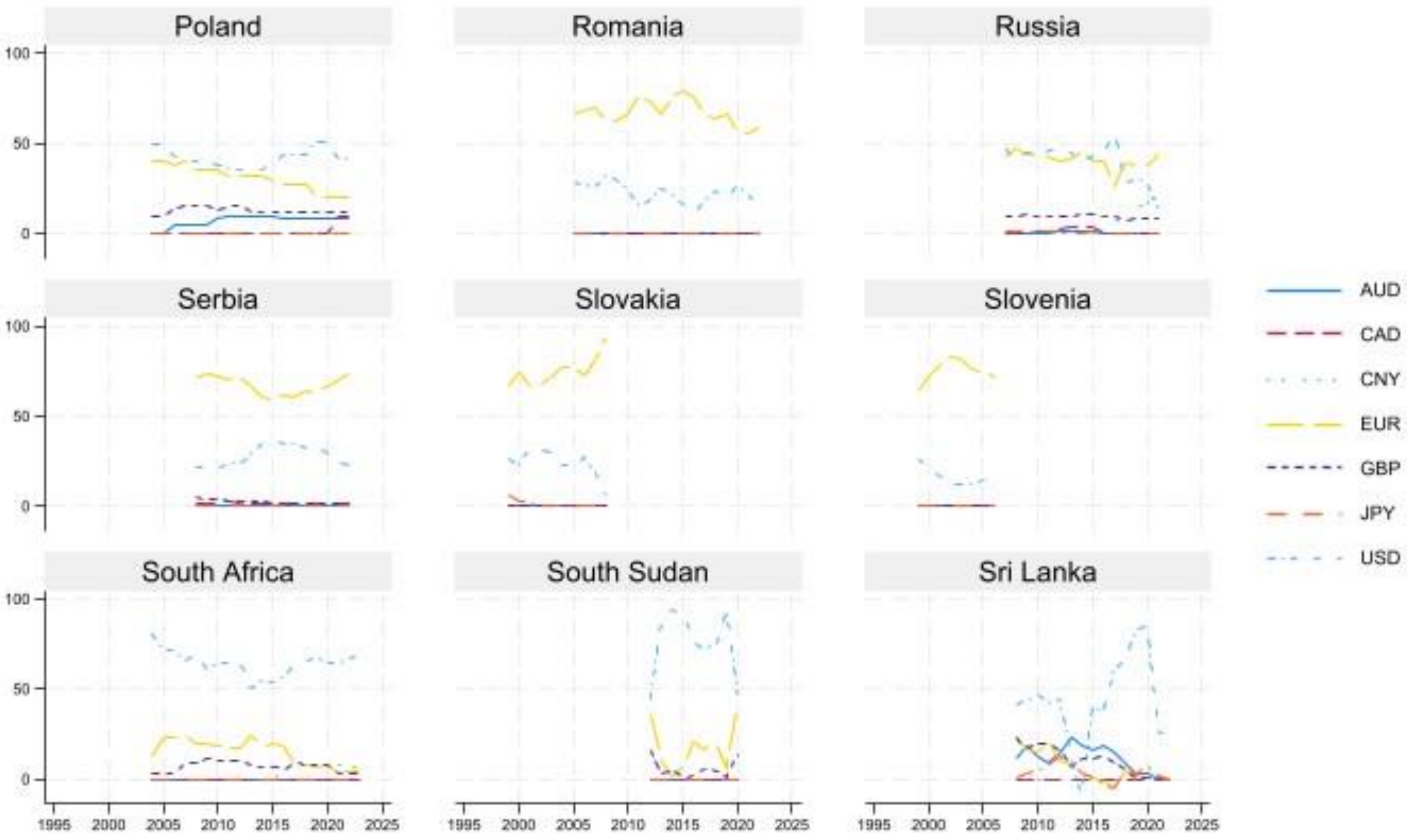

Poland
Romania
Russia
Serbia
Slovakia
Slovenia
South Africa
South Sudan
Sri Lanka
100
50
0
1995
2000
2005
2010
2015
2020
2025
AUD
CAD
CNY
EUR
GBP
JPY
USD

Figure 2.6

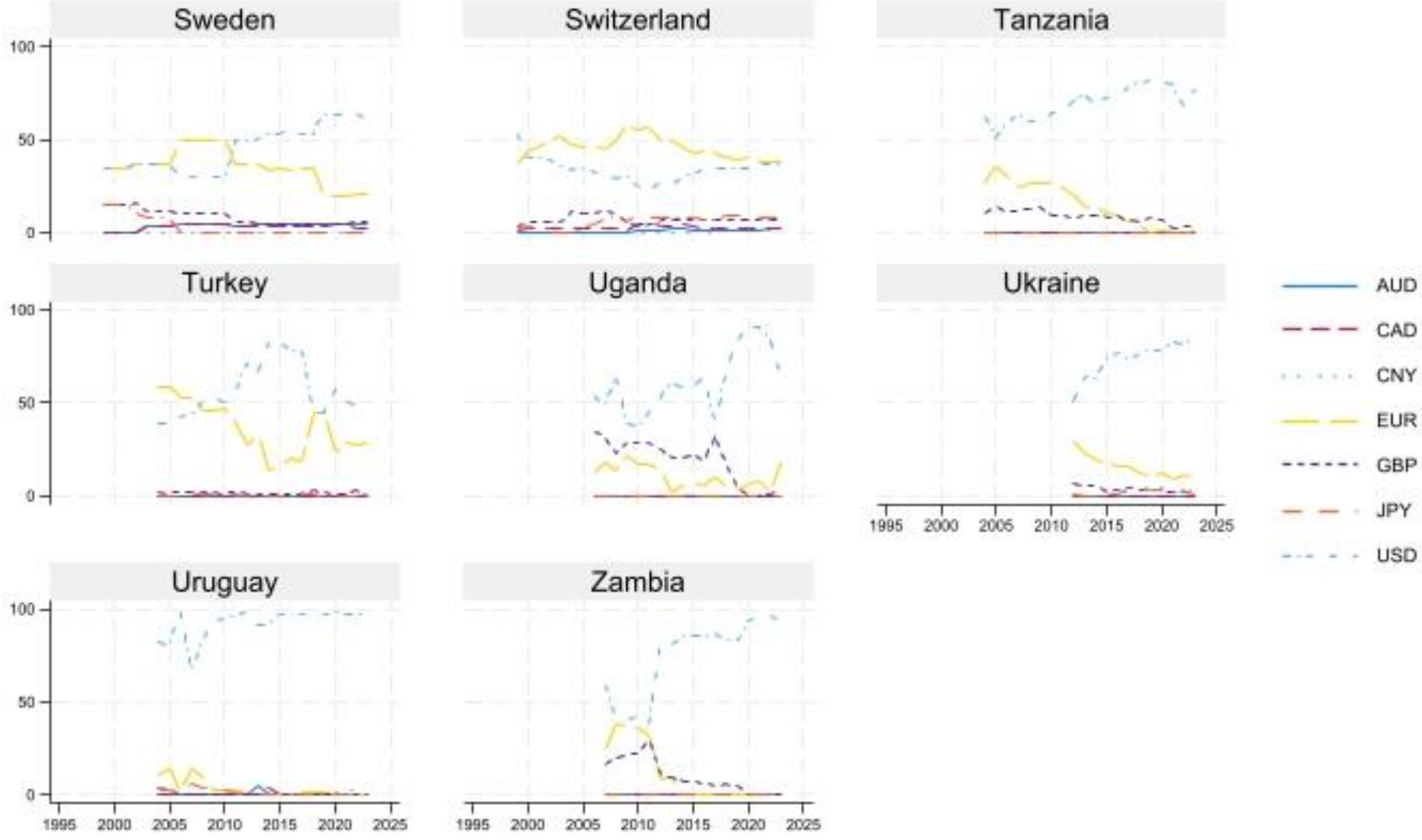

Sweden
Switzerland
Tanzania
Turkey
Uganda
Ukraine
Uruguay
Zambia
AUD
CAD
CNY
EUR
GBP
JPY
USD

## Online Appendix A. Variable Descriptions and Additional Results for Constrained SUR

Table A1. Variable Descriptions

| Variable | Description |
|---|---|
| AUD.L1 | Lagged share of the Australian dollar (AUD) in the previous period. |
| CAD.L1 | Lagged share of the Canadian dollar (CAD) in the previous period. |
| CNY.L1 | Lagged share of the Chinese yuan (CNY) in the previous period. |
| GBP.L1 | Lagged share of the British pound sterling (GBP) in the previous period. |
| JPY.L1 | Lagged share of the Japanese yen (JPY) in the previous period. |
| USD.L1 | Lagged share of the US dollar (USD) in the previous period. |
| US_GDP_share | Share of US GDP in world GDP. |
| EA_GDP_share | Share of euro area GDP in world GDP.<br>The share of China's GDP in world GDP is not included because it is highly collinear with EA_GDP_share ( correlation coefficient = -0.9388). |
| WUI_US | Difference between the US World Uncertainty Index (WUI) and the average WUI of the G7 countries. |
| WUI_EA_big_four | Difference between the average WUI of France, Germany, Italy, and Spain and the average WUI of the G7 countries. (A euro area-wide uncertainty measure is not available.) |
| WUI_CN | Difference between China's WUI and the average WUI of the G7 countries. |
| Trade_Share_with US | A country's share of trade with the United States relative to its total trade with the world. |
| Trade_share_with _EA | A country's share of trade with the euro area relative to its total trade with the world. |
| Trade_share_with_CN | A country's share of trade with China relative to its total trade with the world. |
| INF_diff_US | Difference between US inflation and the average inflation rate across OECD countries. |
| INF_diff_EA | Difference between euro area inflation and the average inflation rate across OECD countries. |
| INF_diff_CN | Difference between China's inflation rate and the average inflation rate across OECD countries. |
| US_SANCTION | Indicator variable equal to 1 if a US sanction is imposed in a given year, and 0 otherwise. |
| EU_SANCTION | Indicator variable equal to 1 if an EU sanction is imposed in a given year, and 0 otherwise. |
| CN_SANCTION | Indicator variable equal to 1 if a Chinese sanction is imposed in a given year, and 0 otherwise. |
| USD_appreciation | Percentage change in the average exchange rate between USD and STR from year $t-1$ to year $t$, where the average exchange rate is the simple average of daily exchange rates.<br>EUR appreciation is not included because it is highly collinear with USD_appreciation (correlation coefficient = -0.8652). |

| Variable | Description |
|---|---|
| CNY_appreciation | Percentage change in the average exchange rate between the CNY and STR from year $t-1$ to year $t$, where the average exchange rate is the simple average of daily exchange rates. |
| USD_volatility | Standard deviation of the daily exchange rate between USD and STR during year $t$. |
| EUR_volatility | Standard deviation of the daily exchange rate between EUR and STR during year $t$. |
| CNY_volatility | Standard deviation of the daily exchange rate between CNY and STR during year $t$. |

Table A2. Currency Composition of Central Bank Reserves (Four-share Specification, One-Way Demeaned)

| ***Variable*** | ***USD*** | ***EUR*** | ***CNY*** | ***OTHER*** |
|---|---|---|---|---|
| *USD.L1* | ***0.6694**** | ***-0.3500**** | *-0.0030* | ***-0.3165**** |
| | *(0.0262)* | *(0.0236)* | *(0.0032)* | *(0.0229)* |
| *CNY.L1* | *-0.1494* | ***-0.4296*** | ***0.8375**** | *-0.2584* |
| | *(0.2053)* | *(0.1854)* | *(0.0253)* | *(0.1800)* |
| *US_GDP_share* | *0.3838* | *-0.2773* | *0.0241* | *-0.1306* |
| | *(0.2364)* | *(0.2135)* | *(0.0291)* | *(0.2072)* |
| *EA_GDP_share* | *-0.1849* | ***0.5974*** | *-0.0075* | ***-0.4051** |
| | *(0.2400)* | *(0.2166)* | *(0.0295)* | *(0.2103)* |
| *WUI_USD* | *-0.0822* | *0.0224* | *-0.0131* | *0.0729* |
| | *(0.0836)* | *(0.0755)* | *(0.0103)* | *(0.0733)* |
| *WUI_Europe* | *0.0822* | ***-0.2270*** | *0.0026* | *0.1421* |
| | *(0.0991)* | *(0.0894)* | *(0.0122)* | *(0.0868)* |
| *WUI_CN* | *-0.0217* | *0.0079* | *-0.0029* | *0.0167* |
| | *(0.0459)* | *(0.0414)* | *(0.0056)* | *(0.0402)* |

| *Variable* | *USD* | *EUR* | *CNY* | *OTHER* |
|---|---|---|---|---|
| *Trade_Share_with US* | *-0.0825* | ***0.3164**** | *-0.0060* | *-0.2279* |
| | *(0.1637)* | *(0.1478)* | *(0.0202)* | *(0.1435)* |
| *Trade_share_with _EA* | *0.0887* | *0.0399* | *-0.0188* | *-0.1097* |
| | *(0.1225)* | *(0.1106)* | *(0.0151)* | *(0.1074)* |
| *Trade_share_with_CN* | ***0.2730***** | ***-0.2717***** | *0.0017* | *-0.0030* |
| | *(0.0771)* | *(0.0696)* | *(0.0095)* | *(0.0676)* |
| *INF_diff_US* | *-0.5677* | *-0.7769* | *0.0474* | ***1.2972**** |
| | *(0.8030)* | *(0.7250)* | *(0.0989)* | *(0.7039)* |
| *INF_diff_EA* | *-0.4731* | *0.8537* | *-0.2265* | *-0.1540* |
| | *(1.4337)* | *(1.2944)* | *(0.1765)* | *(1.2567)* |
| *INF_diff_CN* | *0.1001* | *-0.0412* | *-0.0353* | *-0.0236* |
| | *(0.2775)* | *(0.2505)* | *(0.0342)* | *(0.2432)* |
| *US_SANCTION* | *-0.3143* | ***1.9818**** | ***0.3118***** | ***-1.9793**** |
| | *(1.2561)* | *(1.1341)* | *(0.1547)* | *(1.1010)* |
| *EU_SANCTION* | *-2.2717* | ***-4.4029***** | ***0.5367***** | ***6.1379****** |
| | *(1.8984)* | *(1.7139)* | *(0.2338)* | *(1.6640)* |

| *Variable* | ***USD*** | ***EUR*** | ***CNY*** | ***OTHER*** |
|---|---|---|---|---|
| *CN_SANCTION* | *4.3285* | *-3.7260* | *0.0309* | *-0.6334* |
| | *(3.1956)* | *(2.8850)* | *(0.3935)* | *(2.8010)* |
| *USD_appreciation* | ***0.7050**** | ***-0.8147***** | ***-0.0951***** | *0.2048* |
| | *(0.3685)* | *(0.3327)* | *(0.0454)* | *(0.3230)* |
| *CNY_appreciation* | *-0.2098* | *0.2746* | ***0.0869***** | *-0.1517* |
| | *(0.3405)* | *(0.3074)* | *(0.0419)* | *(0.2985)* |
| *USD_volatility* | ***-2.6956**** | ***2.8443***** | *0.0991* | *-0.2478* |
| | *(1.5012)* | *(1.3554)* | *(0.1849)* | *(1.3159)* |
| *EUR_volatility* | *0.9537* | *-0.6525* | *-0.1079* | *-0.1933* |
| | *(0.9532)* | *(0.8606)* | *(0.1174)* | *(0.8355)* |
| *CNY_volatility* | *0.9853* | *-1.5183* | *-0.3907* | *0.9237* |
| | *(4.5608)* | *(4.1176)* | *(0.5616)* | *(3.9978)* |
| ***Observations*** | *777* | *777* | *777* | *777* |
| ***Adjusted R-squared*** | ***0.5261*** | ***0.4406*** | ***0.6410*** | ***0.1935*** |

*Notes: Standard errors are in parentheses below coefficients. Significant coefficients are shown in bold. * p<0.10, ** p<0.05, *** p<0.001.*

Table A3. Currency Composition of Central Bank Reserves (Four-share Specification, Two-Way Demeaned)

| ***Variable*** | ***USD*** | ***EUR*** | ***CNY*** | ***OTHER*** |
|---|---|---|---|---|
| *USD.L1* | ***0.6794**** | ***-0.3850**** | *-0.0035* | ***-0.2909**** |
| | *(0.0254)* | *(0.0233)* | *(0.0031)* | *(0.0224)* |
| *CNY.L1* | *-0.1215* | ***-0.5066*** | ***0.8380**** | *-0.2099* |
| | *(0.2051)* | *(0.1882)* | *(0.0250)* | *(0.1807)* |
| *US_GDP_share* | *-0.6073* | *0.6182* | *-0.1464* | *0.1354* |
| | *(3.0739)* | *(2.8197)* | *(0.3752)* | *(2.7070)* |
| *EA_GDP_share* | *2.1907* | *-0.8861* | *0.1071* | *-1.4117* |
| | *(2.7899)* | *(2.5592)* | *(0.3406)* | *(2.4569)* |
| *WUI_USD* | *-0.6358* | *0.3735* | *-0.0450* | *0.3074* |
| | *(1.1001)* | *(1.0091)* | *(0.1343)* | *(0.9688)* |
| *WUI_Europe* | *0.6001* | *-0.1811* | *-0.1002* | *-0.3188* |
| | *(1.7056)* | *(1.5645)* | *(0.2082)* | *(1.5020)* |
| *WUI_CN* | *0.6129* | *-0.1168* | *-0.1083* | *-0.3878* |
| | *(0.7827)* | *(0.7180)* | *(0.0955)* | *(0.6893)* |
| *Trade_Share_with US* | *-0.0381* | *0.2333* | *-0.0092* | *-0.1860* |
| | *(0.1609)* | *(0.1476)* | *(0.0196)* | *(0.1417)* |
| *Trade_share_with_EA* | ***-0.1908*** | ***0.6509**** | ***-0.0150** | ***-0.4451**** |
| | *(0.0743)* | *(0.0681)* | *(0.0091)* | *(0.0654)* |
| *Trade_share_with_CN* | ***0.2604*** | ***-0.2644**** | *0.0030* | *0.0010* |
| | *(0.0755)* | *(0.0693)* | *(0.0092)* | *(0.0665)* |
| *INF_diff_US* | *-2.7214* | *1.8652* | *-0.0887* | *0.9448* |
| | *(13.0088)* | *(11.9331)* | *(1.5879)* | *(11.4562)* |
| *INF_diff_EA* | *-21.6715* | *8.2394* | *0.1953* | *13.2368* |
| | *(30.3888)* | *(27.8761)* | *(3.7095)* | *(26.7619)* |

| *Variable* | *USD* | *EUR* | *CNY* | *OTHER* |
|---|---|---|---|---|
| *INF_diff_CN* | *-1.4652* | *1.4939* | *-0.3393* | *0.3106* |
| | *(5.9000)* | *(5.4122)* | *(0.7202)* | *(5.1958)* |
| *US_SANCTION* | *-0.4079* | ***2.5690**** | ***0.3415**** | ***-2.5027**** |
| | *(1.2610)* | *(1.1567)* | *(0.1539)* | *(1.1105)* |
| *EU_SANCTION* | *-2.7659* | ***-4.2832**** | ***0.5170**** | ***6.5321***** |
| | *(1.8928)* | *(1.7363)* | *(0.2310)* | *(1.6669)* |
| *CN_SANCTION* | *3.7251* | *-2.0608* | *0.0212* | *-1.6855* |
| | *(3.2015)* | *(2.9368)* | *(0.3908)* | *(2.8194)* |
| *USD_appreciation* | *0.6066* | *-2.0831* | *0.3395* | *1.1371* |
| | *(5.2476)* | *(4.8137)* | *(0.6406)* | *(4.6213)* |
| *CNY_appreciation* | *-1.2308* | *2.3043* | *-0.3104* | *-0.7630* |
| | *(4.9788)* | *(4.5671)* | *(0.6077)* | *(4.3846)* |
| *USD_volatility* | *-10.8133* | *7.1171* | *0.4807* | *3.2155* |
| | *(16.7187)* | *(15.3363)* | *(2.0408)* | *(14.7233)* |
| *EUR_volatility* | *5.2151* | *-6.3079* | *0.1144* | *0.9784* |
| | *(13.2833)* | *(12.1850)* | *(1.6215)* | *(11.6980)* |
| *CNY_volatility* | *-37.4311* | *32.1535* | *1.8799* | *3.3977* |
| | *(94.6475)* | *(86.8214)* | *(11.5533)* | *(83.3513)* |
| ***Observations*** | ***777*** | ***777*** | ***777*** | ***777*** |
| ***Adjusted R-squared*** | ***0.5376*** | ***0.5230*** | ***0.6197*** | ***0.1912*** |

***Notes:*** *Standard errors are in parentheses below coefficients. Significant coefficients are shown in bold. * p<0.10, ** p<0.05, *** p<0.001*

**Online Appendix B. Modelling Constraints Using Log-ratios**

An alternative approach to imposing the adding-up restriction is to model the log-ratios of reserve currency shares rather than the shares themselves. Expressing each reserve currency share relative to a common benchmark currency, say 'Other', we define

$$z_{ijt} = \ln\left(\frac{\text{share}_{ijt}}{\text{other}_{jt}}\right) \tag{3}$$

Replacing $\text{share}_{ijt}$ and $\text{share}_{ij,t-1}$ in equation (1) with $z_{ijt}$ and $z_{ij,t-1}$, respectively, yields a system of seven log-ratio equations that can be estimated using conventional fixed-effects or random-effects panel estimators. Because the adding-up restriction is incorporated through the transformation itself, no explicit cross-equation constraints are required.

The reserve shares can be recovered from the estimated log-ratios as

$$\text{share}_{ijt} = \frac{\exp(z_{ijt})}{1+\sum_{k=1}^{7}\exp(z_{kjt})} \tag{4}$$

and

$$\text{other}_{jt} = \frac{1}{1+\sum_{k=1}^{7}\exp(z_{kjt})} \tag{5}$$

It follows immediately that

$$\sum_{i=1}^{7}\text{share}_{ijt} + \text{other}_{jt} = 1, \tag{6}$$

so that the adding-up restriction is automatically satisfied.

The log-ratio formulation is an elegant alternative because it is easier to estimate. Its coefficients, however, describe changes relative to the omitted benchmark and must be transformed nonlinearly to obtain effects on portfolio shares. By contrast, the constrained SUR coefficients retain a direct percentage-point interpretation and make the offsetting reallocations across currencies explicit. We therefore adopt constrained SUR in our empirical analyses.

Tables B1 and B2 report the results from models in which the log ratios are computed using 'Other' as the common benchmark. These results are largely consistent with the findings from the constrained SUR models reported in the main body of the paper.

**Table B1. Currency Composition of Central Bank Reserves (Log Ratios, Country Fixed Effects)**

| *Variable* | *AUD* | *CAD* | *CNY* | *EUR* | *GBP* | *JPY* | *USD* |
|---|---|---|---|---|---|---|---|
| *AUD.L1* | ***0.6320**** | ***0.0982**** | *0.0039* | *0.0188* | *-0.0199* | *-0.0005* | *-0.0295* |
| | *(0.0523)* | *(0.0521)* | *(0.0522)* | *(0.0551)* | *(0.0508)* | *(0.0577)* | *(0.0467)* |
| *CAD.L1* | ***0.0985**** | ***0.6850**** | *-0.0461* | *0.0488* | *0.0627* | *-0.0366* | *0.0038* |
| | *(0.0476)* | *(0.0475)* | *(0.0476)* | *(0.0502)* | *(0.0463)* | *(0.0526)* | *(0.0426)* |
| *CNY.L1* | ***0.0816**** | *0.0305* | ***0.8368**** | ***0.1762**** | *0.0612* | *0.0484* | *0.0528* |
| | *(0.0414)* | *(0.0413)* | *(0.0414)* | *(0.0437)* | *(0.0403)* | *(0.0457)* | *(0.0371)* |
| *GBP.L1* | *-0.0749* | *0.0292* | *-0.0168* | *0.0099* | ***0.5846**** | *-0.0854* | *-0.0050* |
| | *(0.0523)* | *(0.0521)* | *(0.0523)* | *(0.0552)* | *(0.0509)* | *(0.0578)* | *(0.0468)* |
| *JPY.L1* | *0.0535* | *0.0071* | *-0.0386* | ***0.0953**** | *0.0339* | ***0.6987**** | *0.0057* |
| | *(0.0387)* | *(0.0385)* | *(0.0386)* | *(0.0408)* | *(0.0376)* | *(0.0427)* | *(0.0346)* |
| *USD.L1* | ***-0.1242**** | ***-0.1932**** | ***-0.1162**** | ***0.2682**** | *-0.0993* | *0.0234* | ***0.6010**** |
| | *(0.0659)* | *(0.0657)* | *(0.0659)* | *(0.0694)* | *(0.0641)* | *(0.0728)* | *(0.0589)* |
| *US_GDP_share* | *-0.1129* | *0.0480* | *0.0343* | *-0.0584* | *-0.0288* | *-0.0089* | *0.0826* |
| | *(0.1591)* | *(0.1586)* | *(0.1591)* | *(0.1677)* | *(0.1548)* | *(0.1757)* | *(0.1424)* |
| *EA_GDP_share* | *-0.0005* | *-0.0045* | *-0.0361* | ***0.2815**** | *0.1424* | *-0.0862* | *-0.0738* |
| | *(0.1555)* | *(0.1549)* | *(0.1554)* | *(0.1639)* | *(0.1513)* | *(0.1717)* | *(0.1391)* |
| *WUI_USD* | *0.0045* | *-0.0020* | *0.0285* | *0.0089* | *0.0363* | *0.0551* | *0.0384* |
| | *(0.0550)* | *(0.0548)* | *(0.0550)* | *(0.0580)* | *(0.0535)* | *(0.0607)* | *(0.0493)* |
| *WUI_Europe* | *0.0308* | *0.0299* | *-0.0426* | *-0.0206* | *-0.0131* | *-0.0625* | *-0.0360* |

| *Variable* | *AUD* | *CAD* | *CNY* | *EUR* | *GBP* | *JPY* | *USD* |
|---|---|---|---|---|---|---|---|
| | (0.0655) | (0.0653) | (0.0655) | (0.0691) | (0.0637) | (0.0723) | (0.0587) |
| WUI_CN | 0.0200 | 0.0307 | -0.0362 | 0.0089 | -0.0116 | -0.0288 | -0.0110 |
| | (0.0301) | (0.0300) | (0.0301) | (0.0317) | (0.0292) | (0.0332) | (0.0269) |
| Trade_Share_with US | -0.0819 | -0.0389 | 0.0016 | 0.1331 | -0.0818 | -0.1413 | 0.0056 |
| | (0.1078) | (0.1074) | (0.1078) | (0.1137) | (0.1049) | (0.1191) | (0.0965) |
| Trade_share_with _EA | -0.0105 | 0.0322 | -0.0199 | 0.0358 | -0.0433 | -0.0732 | 0.0176 |
| | (0.0807) | (0.0804) | (0.0806) | (0.0850) | (0.0785) | (0.0891) | (0.0722) |
| Trade_share_with_CN | 0.0282 | -0.0042 | 0.0010 | **-0.0931*** | 0.0094 | -0.0335 | 0.0067 |
| | (0.0518) | (0.0517) | (0.0518) | (0.0546) | (0.0504) | (0.0572) | (0.0464) |
| INF_diff_US | -0.5442 | -0.4740 | 0.4447 | -0.4535 | 0.1241 | 0.3263 | -0.0392 |
| | (0.5294) | (0.5275) | (0.5291) | (0.5580) | (0.5150) | (0.5845) | (0.4744) |
| INF_diff_EA | -0.3212 | -0.5636 | -0.3543 | -0.4474 | -0.3265 | 0.6687 | -0.0622 |
| | (0.9381) | (0.9346) | (0.9375) | (0.9886) | (0.9124) | (1.0357) | (0.8396) |
| INF_diff_CN | -0.0465 | -0.0628 | **-0.3571*** | -0.0821 | -0.1030 | -0.2125 | -0.0418 |
| | (0.1826) | (0.1820) | (0.1825) | (0.1925) | (0.1776) | (0.2017) | (0.1633) |
| US_SANCTION | -1.2693 | -1.1102 | -0.7982 | -0.8502 | -1.2942 | **-1.7023*** | **-1.2893*** |
| | (0.8141) | (0.8112) | (0.8137) | (0.8580) | (0.7919) | (0.8989) | (0.7281) |
| EU_SANCTION | -0.6979 | **-2.4102*** | 0.0639 | 0.0457 | -0.4619 | -0.7767 | -0.4811 |
| | (1.2569) | (1.2523) | (1.2562) | (1.3247) | (1.2225) | (1.3877) | (1.1241) |
| CN_SANCTION | -0.3553 | -0.6307 | 1.2555 | 0.5677 | 0.3477 | 1.6963 | 1.0701 |
| | (2.0903) | (2.0826) | (2.0891) | (2.2030) | (2.0331) | (2.3078) | (1.8694) |
| USD_appreciation | -0.0425 | -0.1450 | 0.2531 | 0.0068 | 0.1978 | 0.1923 | 0.0881 |
| | (0.2446) | (0.2437) | (0.2445) | (0.2578) | (0.2379) | (0.2701) | (0.2189) |
| CNY_appreciation | 0.1172 | 0.1323 | -0.2833 | 0.1664 | -0.0135 | 0.0129 | 0.0230 |
| | (0.2258) | (0.2250) | (0.2256) | (0.2379) | (0.2196) | (0.2493) | (0.2020) |
| USD_volatility | 0.2139 | 0.2664 | -0.2172 | 0.1717 | 0.4263 | 0.4414 | 0.6360 |
| | (0.9913) | (0.9877) | (0.9908) | (1.0448) | (0.9642) | (1.0945) | (0.8866) |
| EUR_volatility | -0.1328 | -0.1308 | 0.9443 | 0.4687 | 0.2582 | 0.9540 | 0.2585 |
| | (0.6318) | (0.6295) | (0.6314) | (0.6658) | (0.6145) | (0.6975) | (0.5651) |
| CNY_volatility | -2.7942 | -2.7491 | -2.0226 | -4.0598 | -2.7872 | -2.2238 | -1.1376 |
| | (2.9955) | (2.9846) | (2.9938) | (3.1570) | (2.9136) | (3.3073) | (2.6810) |
| **Observations** | **773** | **773** | **773** | **773** | **773** | **773** | **772** |
| **Within R-squared** | **0.4851** | **0.4992** | **0.5276** | **0.3895** | **0.4379** | **0.5028** | **0.4417** |

Notes: Standard errors are in parentheses below coefficients. Significant coefficients are shown in bold. * p<0.10, ** p<0.05, *** p<0.001.

### Table B2. Currency Composition of Central Bank Reserves (Log Ratios, Country and Year Fixed Effects)

| *Variable* | *AUD* | *CAD* | *CNY* | *EUR* | *GBP* | *JPY* | *USD* |
|---|---|---|---|---|---|---|---|
| *AUD.L1* | ***0.6271**** | ***0.1005**** | *-0.0062* | *0.0153* | *-0.0251* | *0.0011* | *-0.0307* |
| | *(0.0531)* | *(0.0530)* | *(0.0532)* | *(0.0561)* | *(0.0516)* | *(0.0586)* | *(0.0475)* |
| *CAD.L1* | ***0.0963**** | ***0.6844**** | *-0.0466* | *0.0479* | *0.0623* | *-0.0372* | *0.0049* |
| | *(0.0478)* | *(0.0478)* | *(0.0479)* | *(0.0505)* | *(0.0465)* | *(0.0528)* | *(0.0428)* |
| *CNY.L1* | *0.0691* | *0.0284* | ***0.8257**** | ***0.1712**** | *0.0510* | *0.0392* | *0.0474* |
| | *(0.0425)* | *(0.0424)* | *(0.0425)* | *(0.0448)* | *(0.0412)* | *(0.0469)* | *(0.0380)* |
| *GBP.L1* | *-0.0691* | *0.0351* | *-0.0143* | *0.0084* | ***0.5908**** | *-0.0953* | *-0.0095* |
| | *(0.0530)* | *(0.0529)* | *(0.0530)* | *(0.0559)* | *(0.0515)* | *(0.0585)* | *(0.0474)* |
| *JPY.L1* | *0.0599* | *0.0067* | *-0.0321* | ***0.0996**** | *0.0388* | ***0.7054**** | *0.0092* |
| | *(0.0391)* | *(0.0390)* | *(0.0391)* | *(0.0413)* | *(0.0380)* | *(0.0432)* | *(0.0350)* |
| *USD.L1* | ***-0.1169**** | ***-0.1978**** | *-0.1040* | ***0.2742**** | *-0.0947* | *0.0301* | ***0.6055**** |
| | *(0.0667)* | *(0.0666)* | *(0.0668)* | *(0.0705)* | *(0.0648)* | *(0.0737)* | *(0.0597)* |
| *US_GDP_share* | *-0.0426* | *0.2055* | *0.0972* | *-0.1314* | *0.1849* | *0.2125* | *0.3662* |
| | *(0.4257)* | *(0.4249)* | *(0.4259)* | *(0.4495)* | *(0.4136)* | *(0.4699)* | *(0.3809)* |
| *EA_GDP_share* | *-0.4186* | *-0.5032* | *-0.4496* | *0.2166* | *0.1686* | *-0.2334* | *-0.2782* |
| | *(0.6869)* | *(0.6857)* | *(0.6872)* | *(0.7253)* | *(0.6673)* | *(0.7582)* | *(0.6145)* |
| *WUI_USD* | *-0.1180* | *0.0195* | *-0.1214* | *-0.1821* | *-0.0937* | *0.0635* | *-0.0178* |
| | *(0.2853)* | *(0.2848)* | *(0.2854)* | *(0.3012)* | *(0.2772)* | *(0.3149)* | *(0.2555)* |
| *WUI_Europe* | *0.0842* | *0.0790* | *0.0458* | *0.0192* | *0.0365* | *0.0031* | *0.0790* |
| | *(0.1809)* | *(0.1805)* | *(0.1809)* | *(0.1910)* | *(0.1757)* | *(0.1996)* | *(0.1618)* |
| *WUI_CN* | *0.0132* | *-0.0263* | *-0.0237* | *0.0348* | *-0.0221* | *-0.1222* | *-0.0516* |
| | *(0.1624)* | *(0.1621)* | *(0.1624)* | *(0.1714)* | *(0.1577)* | *(0.1792)* | *(0.1453)* |
| *Trade_Share_with US* | *-0.0619* | *-0.0347* | *0.0180* | *0.1391* | *-0.0647* | *-0.1462* | *0.0078* |
| | *(0.1092)* | *(0.1090)* | *(0.1092)* | *(0.1153)* | *(0.1060)* | *(0.1205)* | *(0.0977)* |
| *Trade_share_with _EA* | *-0.0052* | *0.0238* | *-0.0146* | *0.0287* | *-0.0396* | *-0.0722* | *0.0133* |
| | *(0.0822)* | *(0.0821)* | *(0.0823)* | *(0.0868)* | *(0.0799)* | *(0.0908)* | *(0.0736)* |
| *Trade_share_with_CN* | *0.0245* | *-0.0058* | *-0.0046* | ***-0.0919**** | *0.0014* | *-0.0257* | *0.0112* |
| | *(0.0525)* | *(0.0524)* | *(0.0526)* | *(0.0555)* | *(0.0510)* | *(0.0580)* | *(0.0470)* |
| *INF_diff_US* | *-1.5766* | *-0.1916* | *-0.4130* | *-1.4873* | *-0.8630* | *-0.0828* | *-0.4888* |
| | *(1.4286)* | *(1.4260)* | *(1.4293)* | *(1.5085)* | *(1.3879)* | *(1.5769)* | *(1.2811)* |
| *INF_diff_EA* | *-0.9280* | *1.8528* | *-0.3913* | *-1.5214* | *-1.7331* | *-0.1020* | *0.4465* |
| | *(4.8041)* | *(4.7955)* | *(4.8064)* | *(5.0727)* | *(4.6671)* | *(5.3028)* | *(4.2978)* |
| *INF_diff_CN* | *-0.0568* | *0.1397* | *-0.3353* | *-0.2560* | *-0.2694* | *-0.3803* | *0.0362* |
| | *(0.7542)* | *(0.7529)* | *(0.7546)* | *(0.7964)* | *(0.7327)* | *(0.8325)* | *(0.6747)* |

| ***Variable*** | ***AUD*** | ***CAD*** | ***CNY*** | ***EUR*** | ***GBP*** | ***JPY*** | ***USD*** |
|---|---|---|---|---|---|---|---|
| *US_SANCTION* | *-1.3079* | *-1.1084* | *-0.8834* | *-0.8555* | ***-1.3649**** | ***-1.6614**** | ***-1.2360**** |
| | *(0.8226)* | *(0.8211)* | *(0.8230)* | *(0.8686)* | *(0.7991)* | *(0.9080)* | *(0.7359)* |
| *EU_SANCTION* | *-0.7714* | ***-2.5341***** | *-0.0322* | *-0.0139* | *-0.5826* | *-0.7878* | *-0.5308* |
| | *(1.2652)* | *(1.2630)* | *(1.2658)* | *(1.3360)* | *(1.2291)* | *(1.3966)* | *(1.1320)* |
| *CN_SANCTION* | *-0.5559* | *-0.7169* | *1.0694* | *0.3530* | *0.2807* | *1.5395* | *0.8788* |
| | *(2.1029)* | *(2.0991)* | *(2.1039)* | *(2.2205)* | *(2.0429)* | *(2.3212)* | *(1.8813)* |
| *USD_appreciation* | *-0.2978* | *-0.3351* | *0.1222* | *0.0181* | *-0.0035* | *-0.3514* | *-0.2644* |
| | *(0.5248)* | *(0.5239)* | *(0.5251)* | *(0.5542)* | *(0.5099)* | *(0.5793)* | *(0.4696)* |
| *CNY_appreciation* | *0.1889* | *0.3806* | *-0.3895* | *-0.1200* | *0.0585* | *0.3395* | *0.1685* |
| | *(0.6874)* | *(0.6862)* | *(0.6877)* | *(0.7258)* | *(0.6678)* | *(0.7588)* | *(0.6150)* |
| *USD_volatility* | *-1.3505* | *0.3651* | *-2.7165* | *-2.7907* | *-1.0255* | *1.1361* | *-0.1488* |
| | *(5.1976)* | *(5.1883)* | *(5.2001)* | *(5.4883)* | *(5.0494)* | *(5.7372)* | *(4.6514)* |
| *EUR_volatility* | *0.7081* | *0.4842* | *2.0257* | *1.7606* | *0.7608* | *1.0327* | *0.7214* |
| | *(1.9663)* | *(1.9627)* | *(1.9672)* | *(2.0762)* | *(1.9102)* | *(2.1704)* | *(1.7591)* |
| *CNY_volatility* | *6.3441* | *2.1421* | *8.8571* | *8.1527* | *6.3056* | *3.1489* | *6.9288* |
| | *(11.1794)* | *(11.1594)* | *(11.1848)* | *(11.8045)* | *(10.8606)* | *(12.3400)* | *(10.0013)* |
| ***Observations*** | ***773*** | ***773*** | ***773*** | ***773*** | ***773*** | ***773*** | ***772*** |
| ***Within R-squared*** | ***0.4900*** | ***0.5021*** | ***0.5311*** | ***0.3930*** | ***0.4446*** | ***0.5078*** | ***0.4466*** |

*Notes: Standard errors are in parentheses below coefficients. Significant coefficients are shown in bold. * p<0.10, ** p<0.05, *** p<0.001.*